\documentclass[conference,a4paper]{IEEEtran}
\usepackage[dvips]{graphicx,color}
\usepackage{latexsym}
\usepackage{amssymb}
\usepackage{array}

\begin{document}

%
%
\newcommand{\qed}{\hfill$\square$}
\newcommand{\suchthat}{\mbox{~s.t.~}}
\newcommand{\markov}{\leftrightarrow}
\newenvironment{pRoof}{%
 \noindent{\em Proof.\ }}{%
 \hspace*{\fill}\qed \\
 \vspace{2ex}}


\newcommand{\ket}[1]{| #1 \rangle}
\newcommand{\bra}[1]{\langle #1 |}
\newcommand{\bol}[1]{\mathbf{#1}}
\newcommand{\rom}[1]{\mathrm{#1}}
\newcommand{\san}[1]{\mathsf{#1}}
\newcommand{\mymid}{:~}
\newcommand{\argmax}{\mathop{\rm argmax}\limits}
\newcommand{\argmin}{\mathop{\rm argmin}\limits}

\newcommand{\Cls}{class NL}
\newcommand{\vSpa}{\vspace{0.3mm}}
\newcommand{\Prmt}{\zeta}
\newcommand{\pj}{\omega_n}

\newfont{\bg}{cmr10 scaled \magstep4}
\newcommand{\bigzerol}{\smash{\hbox{\bg 0}}}
\newcommand{\bigzerou}{\smash{\lower1.7ex\hbox{\bg 0}}}
\newcommand{\nbn}{\frac{1}{n}}
\newcommand{\ra}{\rightarrow}
\newcommand{\la}{\leftarrow}
\newcommand{\ldo}{\ldots}
\newcommand{\typi}{A_{\epsilon }^{n}}
\newcommand{\bx}{\hspace*{\fill}$\Box$}
\newcommand{\pa}{\vert}
\newcommand{\ignore}[1]{}

%
%
%
%
\newcommand{\bc}{\begin{center}}  %
\newcommand{\ec}{\end{center}}
\newcommand{\befi}{\begin{figure}[h]}  %
\newcommand{\enfi}{\end{figure}}
\newcommand{\bsb}{\begin{shadebox}\begin{center}}   %
\newcommand{\esb}{\end{center}\end{shadebox}}
\newcommand{\bs}{\begin{screen}}     %
\newcommand{\es}{\end{screen}}
\newcommand{\bib}{\begin{itembox}}   %
\newcommand{\eib}{\end{itembox}}
\newcommand{\bit}{\begin{itemize}}   %
\newcommand{\eit}{\end{itemize}}
\newcommand{\defeq}{\stackrel{\triangle}{=}}
\newcommand{\Qed}{\hbox{\rule[-2pt]{3pt}{6pt}}}
\newcommand{\beq}{\begin{equation}}
\newcommand{\eeq}{\end{equation}}
\newcommand{\beqa}{\begin{eqnarray}}
\newcommand{\eeqa}{\end{eqnarray}}
\newcommand{\beqno}{\begin{eqnarray*}}
\newcommand{\eeqno}{\end{eqnarray*}}
\newcommand{\ba}{\begin{array}}
\newcommand{\ea}{\end{array}}
\newcommand{\vc}[1]{\mbox{\boldmath $#1$}}
\newcommand{\lvc}[1]{\mbox{\scriptsize \boldmath $#1$}}
\newcommand{\svc}[1]{\mbox{\scriptsize\boldmath $#1$}}

\newcommand{\wh}{\widehat}
\newcommand{\wt}{\widetilde}
\newcommand{\ts}{\textstyle}
\newcommand{\ds}{\displaystyle}
\newcommand{\scs}{\scriptstyle}
\newcommand{\vep}{\varepsilon}
\newcommand{\rhp}{\rightharpoonup}
\newcommand{\cl}{\circ\!\!\!\!\!-}
\newcommand{\bcs}{\dot{\,}.\dot{\,}}
\newcommand{\eqv}{\Leftrightarrow}
\newcommand{\leqv}{\Longleftrightarrow}
\newtheorem{co}{Corollary} 
\newtheorem{lm}{Lemma} 
\newtheorem{Ex}{Example} 
\newtheorem{Th}{Theorem}
\newtheorem{df}{Definition} 
\newtheorem{pr}{Property} 
\newtheorem{pro}{Proposition} 
\newtheorem{rem}{Remark} 

\newcommand{\auxan}{{X}^n}
\newcommand{\auxbn}{{U}^n}
\newcommand{\rvxn}{\empty}
\newcommand{\vauxa}{{\vc X}}
\newcommand{\vauxb}{{\vc U}}
\newcommand{\vx}{{\empty}}
\newcommand{\lvauxa}{{\lvc X}}
\newcommand{\lvauxb}{{\lvc U}}
\newcommand{\lvx}{{\empty}}

\newcommand{\lcv}{convex } 

\newcommand{\hugel}{{\arraycolsep 0mm
                    \left\{\ba{l}{\,}\\{\,}\ea\right.\!\!}}
\newcommand{\Hugel}{{\arraycolsep 0mm
                    \left\{\ba{l}{\,}\\{\,}\\{\,}\ea\right.\!\!}}
\newcommand{\HUgel}{{\arraycolsep 0mm
                    \left\{\ba{l}{\,}\\{\,}\\{\,}\vspace{-1mm}
                    \\{\,}\ea\right.\!\!}}
\newcommand{\huger}{{\arraycolsep 0mm
                    \left.\ba{l}{\,}\\{\,}\ea\!\!\right\}}}

\newcommand{\Huger}{{\arraycolsep 0mm
                    \left.\ba{l}{\,}\\{\,}\\{\,}\ea\!\!\right\}}}

\newcommand{\HUger}{{\arraycolsep 0mm
                    \left.\ba{l}{\,}\\{\,}\\{\,}\vspace{-1mm}
                    \\{\,}\ea\!\!\right\}}}

\newcommand{\hugebl}{{\arraycolsep 0mm
                    \left[\ba{l}{\,}\\{\,}\ea\right.\!\!}}
\newcommand{\Hugebl}{{\arraycolsep 0mm
                    \left[\ba{l}{\,}\\{\,}\\{\,}\ea\right.\!\!}}
\newcommand{\HUgebl}{{\arraycolsep 0mm
                    \left[\ba{l}{\,}\\{\,}\\{\,}\vspace{-1mm}
                    \\{\,}\ea\right.\!\!}}
\newcommand{\hugebr}{{\arraycolsep 0mm
                    \left.\ba{l}{\,}\\{\,}\ea\!\!\right]}}
\newcommand{\Hugebr}{{\arraycolsep 0mm
                    \left.\ba{l}{\,}\\{\,}\\{\,}\ea\!\!\right]}}

\newcommand{\HugebrB}{{\arraycolsep 0mm
                    \left.\ba{l}{\,}\\{\,}\vspace*{-1mm}\\{\,}\ea\!\!\right]}}

\newcommand{\HUgebr}{{\arraycolsep 0mm
                    \left.\ba{l}{\,}\\{\,}\\{\,}\vspace{-1mm}
                    \\{\,}\ea\!\!\right]}}

\newcommand{\hugecl}{{\arraycolsep 0mm
                    \left(\ba{l}{\,}\\{\,}\ea\right.\!\!}}
\newcommand{\Hugecl}{{\arraycolsep 0mm
                    \left(\ba{l}{\,}\\{\,}\\{\,}\ea\right.\!\!}}
\newcommand{\hugecr}{{\arraycolsep 0mm
                    \left.\ba{l}{\,}\\{\,}\ea\!\!\right)}}
\newcommand{\Hugecr}{{\arraycolsep 0mm
                    \left.\ba{l}{\,}\\{\,}\\{\,}\ea\!\!\right)}}

\newcommand{\hugepl}{{\arraycolsep 0mm
                    \left|\ba{l}{\,}\\{\,}\ea\right.\!\!}}
\newcommand{\Hugepl}{{\arraycolsep 0mm
                    \left|\ba{l}{\,}\\{\,}\\{\,}\ea\right.\!\!}}
\newcommand{\hugepr}{{\arraycolsep 0mm
                    \left.\ba{l}{\,}\\{\,}\ea\!\!\right|}}
\newcommand{\Hugepr}{{\arraycolsep 0mm
                    \left.\ba{l}{\,}\\{\,}\\{\,}\ea\!\!\right|}}

\newcommand{\MEq}[1]{\stackrel{
{\rm (#1)}}{=}}

\newcommand{\MLeq}[1]{\stackrel{
{\rm (#1)}}{\leq}}

\newcommand{\ML}[1]{\stackrel{
{\rm (#1)}}{<}}

\newcommand{\MGeq}[1]{\stackrel{
{\rm (#1)}}{\geq}}

\newcommand{\MG}[1]{\stackrel{
{\rm (#1)}}{>}}

\newcommand{\MPreq}[1]{\stackrel{
{\rm (#1)}}{\preceq}}

\newcommand{\MSueq}[1]{\stackrel{
{\rm (#1)}}{\succeq}}

\newenvironment{jenumerate}
	{\begin{enumerate}\itemsep=-0.25em \parindent=1zw}{\end{enumerate}}
\newenvironment{jdescription}
	{\begin{description}\itemsep=-0.25em \parindent=1zw}{\end{description}}
\newenvironment{jitemize}
	{\begin{itemize}\itemsep=-0.25em \parindent=1zw}{\end{itemize}}
\renewcommand{\labelitemii}{$\cdot$}

\newcommand{\iro}[2]{{\color[named]{#1}#2\normalcolor}}
\newcommand{\irr}[1]{{\color[named]{Red}#1\normalcolor}}
\newcommand{\irg}[1]{{\color[named]{Green}#1\normalcolor}}
\newcommand{\irb}[1]{{\color[named]{Blue}#1\normalcolor}}
\newcommand{\irBl}[1]{{\color[named]{Black}#1\normalcolor}}
\newcommand{\irWh}[1]{{\color[named]{White}#1\normalcolor}}

\newcommand{\irY}[1]{{\color[named]{Yellow}#1\normalcolor}}
\newcommand{\irO}[1]{{\color[named]{Orange}#1\normalcolor}}
\newcommand{\irBr}[1]{{\color[named]{Purple}#1\normalcolor}}
\newcommand{\IrBr}[1]{{\color[named]{Purple}#1\normalcolor}}
\newcommand{\irBw}[1]{{\color[named]{Brown}#1\normalcolor}}
\newcommand{\irPk}[1]{{\color[named]{Magenta}#1\normalcolor}}
\newcommand{\irCb}[1]{{\color[named]{CadetBlue}#1\normalcolor}}

%
\newenvironment{indention}[1]{\par
\addtolength{\leftskip}{#1}\begingroup}{\endgroup\par}
%
\newcommand{\namelistlabel}[1]{\mbox{#1}\hfill} 
\newenvironment{namelist}[1]{%
\begin{list}{}
{\let\makelabel\namelistlabel
\settowidth{\labelwidth}{#1}
\setlength{\leftmargin}{1.1\labelwidth}}
}{%
\end{list}}
%
%
\newcommand{\bfig}{\begin{figure}[t]}
\newcommand{\efig}{\end{figure}}
\setcounter{page}{1}

\newtheorem{theorem}{Theorem}

\newcommand{\ep}{\mbox{\rm e}}

\newcommand{\Exp}{\exp
}
\newcommand{\idenc}{{\varphi}_n}
\newcommand{\resenc}{
{\varphi}_n}
\newcommand{\ID}{\mbox{\scriptsize ID}}
\newcommand{\TR}{\mbox{\scriptsize TR}}
\newcommand{\Av}{\mbox{\sf E}}

\newcommand{\Vl}{|}
\newcommand{\Ag}{(R,P_{X^n}|W^n)}
\newcommand{\Agv}[1]{({#1},P_{X^n}|W^n)}
\newcommand{\Avw}[1]{({#1}|W^n)}

\newcommand{\Jd}{X^nY^n}
\newcommand{\IdR}{r_n}

\newcommand{\Index}{{n,i}}

\newcommand{\cid}{C_{\mbox{\scriptsize ID}}}
\newcommand{\cida}{C_{\mbox{{\scriptsize ID,a}}}}

\newcommand{\iN}{\rm (in)}
\newcommand{\ouT}{\rm (out)}

\arraycolsep 0.5mm
\date{}
%
\title{
Strong Converse Exponent for Degraded Broadcast Channels 
at Rates outside the Capacity Region
}
\author{%
\IEEEauthorblockA{
Yasutada Oohama\\
  University of Electro-Communications, Tokyo, Japan \\
  Email: oohama@uec.ac.jp} 
} 

\maketitle


\begin{abstract} 
We consider the discrete memoryless degraded broadcast channels. 
We prove that the error probability of decoding tends to one 
exponentially for rates outside the capacity region and 
derive an explicit lower bound of this exponent function. 
We shall demonstrate that the information spectrum approach 
is quite useful for investigating this problem.
\end{abstract}
%
\section{The Capacity Region of the Degraded Broadcast Channels}

\begin{figure}[t]
\bc
\includegraphics[width=7.6cm]{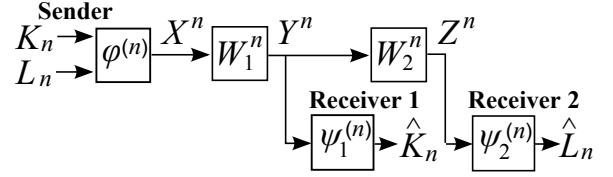}
\caption{Transmission of messages via the degraded BC.}
\label{fig:theGBC} 
\ec
\end{figure}

Let ${\cal X}, {\cal Y},$ ${\cal Z}$ be finite sets.
The broadcast channel we study in this paper is defined 
by a discrete memoryless channel specified with the following 
stochastic matrix:
\beq
{W} \defeq \{ {W}(y,z|x)\}_{(x,y,z) 
\in    {\cal X}
\times {\cal Y} 
\times {\cal Z}}.
\eeq
Here the set ${\cal X}$ stands for a set of channel input. 
The sets ${\cal Y}$ and ${\cal Z}$ stand for sets of 
two channel outputs. Let $X^n$ be a random 
variable taking values in ${\cal X}^n$. 
We write an element of ${\cal X}^n$ as   
$x^n=x_{1}x_{2}$$\cdots x_{n}.$ 
Suppose that $X^n$ has a probability distribution on ${\cal X}^n$ 
denoted by 
$p_{X^n}=$ 
$\left\{p_{X^n}(x^n) \right\}_{{x^n} \in {\cal X}^n}$.
Similar notations are adopted for other random variables. 
Let $Y^n \in {\cal Y}^n$ and $Z^n \in {\cal Y}^n$  be random variables 
obtained as the channel output by connecting 
$X^n$ to the input of channel. We write 
a conditional distribution of $(Y^n,Z^n)$ on given $X^n$ 
as 
$$
W^n=
\left\{W^n(y^n,z^n|x^n)\right
\}_{(x^n,y^n,z^n)\in {\cal X}^n \times {\cal Y}^n \times {\cal Z}^n}.
$$
In this paper we deal with the case where the components 
$W({z},{y}|{x})$ of $W$ satisfy 
the following conditions:
\beq
W({y},{z}|{x})=W_1({ y}|{ x})W_2({ z}|{ y}).
\label{eqn:sde1}
\eeq
In this case we say that the broadcast channel ${W}$ 
is {\it degraded}. The degraded broadcast channel (DBC) 
is specified by $(W_1,W_2)$. 
Transmission of messages via the degraded BC is shown 
in Fig. \ref{fig:theGBC}. Let $K_n$ and  $L_n$ be uniformly 
distributed random variables taking values in message sets 
${\cal K}_n $ and ${\cal L}_n$, respectively. 
The random variable $K_n$ is a message sent to the receiver 1.
The random variable $L_n$ is a message sent to the receiver 2.
A sender transforms $K_n$ and $L_n$ into a transmitted 
sequence $X^n$ using an encoder function $\varphi^{(n)}$ and 
sends it to the receivers 1 and 2. 
In this paper we assume that the encoder function $\varphi^{(n)}$ 
is a stochastic encoder. In this case, $\varphi^{(n)}$ is 
a stochastic matrix given by
$$
\varphi^{(n)}=\{
\varphi^{(n)}(x^n|k,l)\}_{
(k,l,x^n)\in {\cal K}_n\times {\cal L}_n \times {\cal X}^n},
$$ 
where $\varphi^{(n)}(x^n|k,l)$ is a conditional probability 
of $x^n \in {\cal X}^n$ given message pair $(k,l)\in$
${\cal K}_n\times {\cal L}_n$.
The joint probability mass function on 
${\cal K}_n \times {\cal L}_n$ 
$\times {\cal X}^n$ 
$\times {\cal Y}^n$ 
$\times {\cal Z}^n$ 
is given by
\beqno
& &\Pr\{(K_n,L_n,X^n,Y^n,Z^n)=(k,l,x^n,y^n, z^n)\}
\nonumber\\
&=&
\frac{\varphi^{(n)}(x^n|k,l)}{\pa{\cal K}_n\pa \pa{\cal L}_n\pa}
\prod_{t=1}^n W_1\left(y_t\left|x_t\right.\right)
              W_2\left(z_t\left|y_t\right.\right),
\eeqno
where $\pa {\cal K}_n \pa$ is a cardinality 
of the set ${\cal K}_n$. The decoding functions 
at the receiver 1 and the receiver 2, respectively, 
are denoted by ${\psi}_1^{(n)}$ and ${\psi}_2^{(n)}$. 
Those functions are formally defined by
$
{\psi}_1^{(n)}: {\cal Y}^{n} \to {\cal K}_n,
{\psi}_2^{(n)}: {\cal Z}^{n} \to {\cal L}_n.
$
The average error probabilities of decoding at the receivers 1 
and 2 are defined by
\beqno
& &{\rm P}_{\rm e,1}^{(n)}={\rm P}_{\rm e}^{(n)}(\varphi^{(n)},\psi_1^{(n)})
 \defeq \Pr\{\psi_{1}^{(n)}(Y^n)\neq K_n\},
\\
& &{\rm P}_{\rm e,2}^{(n)}={\rm P}_{\rm e}^{(n)}(\varphi^{(n)},\psi_2^{(n)})
\defeq \Pr\{\psi_{2}^{(n)}(Z^n)\neq L_n\}.
\eeqno
Furthermore, we set
\beqno
& &{\rm P}_{\rm e}^{(n)}
={\rm P}_{\rm e}^{(n)}(\varphi^{(n)},\psi_1^{(n)},\psi_2^{(n)})
\\
& \defeq & \Pr\{\psi_{1}^{(n)}(Y^n)\neq K_n \mbox{ or } 
\psi_{2}^{(n)}(Z^n)\neq L_n \}.
\eeqno
It is obvious that we have the following relation.
\beq
{\rm P}_{\rm e}^{(n)}\leq 
{\rm P}_{\rm e,1}^{(n)}+ {\rm P}_{\rm e,2}^{(n)}. 
\eeq%
For $k\in {\cal K}_n$ and $l\in {\cal L}_n$, set
$
{\cal D}_1(k)\defeq$ $\{ y^n: \psi_1^{(n)}(y^n)=k \},
$
$
{\cal D}_2(l)\defeq$ $\{ z^n: \psi_2^{(n)}(z^n)=l \}.
$
The families of sets 
$\{ {\cal D}_1(k) \}_{k\in {\cal K}_n}$ and
$\{ {\cal D}_2(l) \}_{l \in {\cal L}_n}$ are called 
the decoding regions. Using the decoding region, 
${\rm P}_{\rm e}^{(n)}$ can be written as
\beqno
{\rm P}_{\rm e}^{(n)}
&=&\frac{1}{|{\cal K}_n| |{\cal L}_n|} 
\sum_{(k,l)\in {\cal K}_n \times {\cal L}_n }
\sum_{\scs (x^n,y^n,z^n)\in {\cal X}^n \times {\cal Y}^n\times {\cal Z}^n:
       \atop{
       \scs y^n   \in {\cal D}_1^{c}(k)\mbox{ or }  
           \scs z^n  \in  {\cal D}_2^{c}(l)
       }
    } 1%
\\
& &\times 
\varphi^{(n)}(x^n|k,l)W_1^n(y^n|x^n)W_2^n(z^n|y^n).
\eeqno
Set 
\beqno
& &{\rm P}^{(n)}_{\rm c}=
   {\rm P}^{(n)}_{\rm c}(\varphi^{(n)},\psi_1^{(n)},\psi_2^{(n)})
\defeq 
1-{\rm P}^{(n)}_{\rm e}(\varphi^{(n)}, \psi_1^{(n)},\psi_2^{(n)}).
\eeqno
The quantity ${\rm P}^{(n)}_{\rm c}$ is called the average 
correct probability of decoding. 
\newcommand{\ZapZz}{
This quantity has the following form
\beqno
{\rm P}_{\rm c}^{(n)}
&=&\frac{1}{|{\cal K}_n| |{\cal L}_n|} 
\sum_{(k,l)\in {\cal K}_n \times {\cal L}_n }
\sum_{\scs (x^n,y^n,z^n)\in {\cal X}^n \times {\cal Y}^n\times {\cal Z}^n:
       \atop{
       \scs y^n   \in {\cal D}_1(k), 
           \scs z^n  \in  {\cal D}_2(l)
       }
    }  
\\
& &\times 
\varphi^{(n)}(x^n|k,l)W_1^n(y^n|x^n)W_2^n(z^n|y^n).
\eeqno
}
For given $(\varepsilon_1,\varepsilon_2)$ $\in (0,1)^2$, 
a pair $(R_1,R_2)$ is $(\varepsilon_1,\varepsilon_2)$-{\it achievable} 
if there exists 
a sequence of triples 
$\{(\varphi^{(n)},$ $\psi_1^{(n)},\psi_2^{(n)})$ $\}_{n=1}^{\infty}$ 
such that 
\beqa 
{\rm P}_{{\rm e,}i}^{(n)}(\varphi^{(n)},\psi_i^{(n)})
 & \leq & \varepsilon_i, i=1,2, 
\nonumber\\
\liminf_{n\to\infty} \nbn \log \pa {\cal K}_n \pa & \geq & R_1,
\liminf_{n\to\infty} \
\nbn \log \pa {\cal L}_n \pa \geq R_2.
\nonumber
\eeqa
The set that consists of all $(\varepsilon_1,\varepsilon_2)$-achievable 
rate pair is denoted by 
${\cal C}_{\rm DBC}(\varepsilon_1,\varepsilon_2| W_1,W_2)$, 
which is called the capacity region of the DBC.
We can define another capacity region based on the error probability 
${\rm P}_{{\rm e}}^{(n)}(\varphi^{(n)},$ $\psi_1^{(n)},$ $\psi_2^{(n)})$.
For given $\varepsilon$ $\in (0,1)$, a pair $(R_1,R_2)$ 
is $\varepsilon$-{\it achievable} 
if there exists a sequence of triples 
$\{(\varphi^{(n)},$ 
$\psi_1^{(n)}, \psi_2^{(n)})\}_{n=1}^{\infty}$ 
such that 
\beqa 
&&{\rm P}_{{\rm e}}^{(n)}
(\varphi^{(n)},\psi_1^{(n)},\psi_2^{(n)})
\leq \varepsilon, 
\nonumber\\
& &\liminf_{n\to\infty} \nbn \log \pa {\cal K}_n \pa  \geq R_1,
\liminf_{n\to\infty} \
\nbn \log \pa {\cal L}_n \pa \geq R_2.
\nonumber
\eeqa
The set that consists of all $\varepsilon$-achievable rate pair is denoted by 
${\cal C}_{\rm DBC}(\varepsilon|W_1,W_2)$. It is obvious 
that for $0< \varepsilon_1+\varepsilon_2\leq 1$, we have
$$
{\cal C}_{\rm DBC}(\varepsilon_1,\varepsilon_2|W_1,W_2)
\subseteq {\cal C}_{\rm DBC}(\varepsilon_1+\varepsilon_2|W_1,W_2).
$$
We set
$$
{\cal C}_{\rm DBC}(W_1,W_2)
\defeq \bigcap_{\varepsilon\in(0,1)}
{\cal C}_{\rm DBC}(\varepsilon|W_1,W_2),
$$
which is called the capacity region of the DBC. 
The two maximum error probabilities of decoding 
are defined by as follows:
\beqno
{\rm P}_{{\rm e,{m}},1}^{(n)}&=&
   {\rm P}_{{\rm e,{m},1}}^{(n)}(\varphi^{(n)},\psi_1^{(n)})
\\
& \defeq & \max_{(k,l) \in {\cal K}_n\times {\cal L}_n}
\Pr\{\psi_1^{(n)}(Y^n)\neq k|K_n=k\},
\\
{\rm P}_{{\rm e,{m}},2}^{(n)}&=&
   {\rm P}_{\rm e,{m},2}^{(n)}(\varphi^{(n)},\psi_2^{(n)})
\\
& \defeq & \max_{l \in {\cal L}_n}
\Pr\{\psi_2^{(n)}(Z^n)\neq l|L_n=l\}.
\eeqno
Based on those quantities, we 
define the {maximum capacity region} 
${\cal C}_{\rm m, DBC}
(\varepsilon_1,\varepsilon_2|{W_1},{W_2})$
in a manner quite similar to the definition of 
${\cal C}_{\rm DBC} 
(\varepsilon_1,\varepsilon_2|{W_1},{W_2})$.
\newcommand{\Zass}{
as follows. 
For a given $(\varepsilon_1,\varepsilon_2) \in (0,1)^2$, 
a pair $(\irb{R_1},\irb{R_2})$ 
is $(\varepsilon_1,\varepsilon_2)$-{\it achievable} 
if there exists a sequence of triples 
$\{(\varphi^{(n)},$ $\psi_1^{(n)}, \psi_2^{(n)})\}_{n=1}^{\infty}$ 
such that
\vspace*{-2mm}
\beqa 
{\rm P}_{{\rm e},{\rm m},i}^{(n)}
(\varphi^{(n)},\psi_i^{(n)})
&\leq &\varepsilon_i,i=1,2, 
\nonumber\\
\liminf_{n\to\infty} \nbn \log \pa {\cal K}_n \pa & \geq & \irb{R_1},
\liminf_{n\to\infty} \
\nbn \log \pa {\cal L}_n \pa \geq \irb{R_{2}}.
\nonumber
\eeqa
The set that consists of all 
$(\varepsilon_1,\varepsilon_2)$-achievable 
rate pair is denoted by ${\cal C}_{{\rm m}, \rm DBC}
(\varepsilon_1,\varepsilon_2|\irBr{W_1},\irBr{W_2})$.
}
To describe previous works on 
${\cal C}_{\rm DBC}(W_1,W_2)$ and 
${\cal C}_{\rm m, DBC}(\varepsilon_1,\varepsilon_2|W_1,W_2)$,
we introduce an auxiliary random variable $U$ taking values in a finite 
set ${\cal U}$. We assume that the joint distribution of 
$(U,X,Y,Z)$ is 
$$ 
p_{U{X}{Y}Z}(u,x,y,z)=p_{U}(u)p_{X|U}(x|u)
W_1(y|x)W_2(z|y). 
$$
The above condition is equivalent to $U \markov 
X \leftrightarrow Y $ $\markov Z$. Define the set of probability 
distribution $p=p_{UXYZ}$  of $(U,$ $X,$ $Y,$ $Z)$ $\in$ ${\cal U}$
$\times{\cal X}$ $\times{\cal Y}$ $\times{\cal Z}$
by
\beqno
&&{\cal P}(W_1,W_2)
\defeq 
\{p: \pa {\cal U} \pa 
\leq \pa {\cal X} \pa + 1,
\vSpa\\
& &\quad p_{Y|X}=W_1, p_{Z|Y}=W_2, 
U \markov  X\markov Y \markov Z \}.
\eeqno
Set 
\beqno
{\cal C}(p)
&\defeq &
\ba[t]{l}
\{(R_1,R_2): R_1,R_2 \geq 0\,,
\vSpa\\
\ba{rcl}
R_1 & \leq & I_p(X;Y|U), 
R_2  \leq I_p(U;Z)\}.
\ea
\ea
\\
{\cal C}(W_1,W_2)
&=& \bigcup_{p\in {\cal P}(W_1,W_2)}
{\cal C}(p).
\eeqno
%
%
\newcommand{\ApdaRegAAA}{
}{
We can show that the above functions and sets 
satisfy the following property. 
\begin{pr}\label{pr:pro0}  
$\quad$
\begin{itemize}
\item[a)] The region ${\cal C}(W_1,W_2)$ is a closed convex set of 
The region ${\cal C}(W_1,W_2)$ is a closed convex 
subset of $\mathbb{R}_{+}^2$, where
\beqno
\mathbb{R}_{+}^2&\defeq &\{(R_1,R_2): R_1 \geq 0,R_2 \geq 0\}.
\eeqno

\item[b)] The region ${\cal C}(W_1,W_2)$ can be expressed 
with a family of supporting hyperplanes. To describe this result 
we define the set of probability distribution 
$p=p_{UXYZ}$ of $(U,$ $X,$ $Y,$ $Z)$ $\in$ ${\cal U}$
$\times{\cal X}$ $\times{\cal Y}$ $\times{\cal Z}$ by
\beqno
& &{\cal P}_{\rm sh}(W_1,W_2)
\defeq 
\{p: 
\pa {\cal U} \pa \leq \pa {\cal X} \pa, 
\vSpa\\
& &\quad p_{Y|X}=W_1, p_{Z|Y}=W_2, 
U \markov  X\markov Y \markov Z \}.
\eeqno
We set
\beqno
& &C^{(\mu)}(W_1,W_2) 
\\
&\defeq& 
\max_{p \in {\cal P}_{\rm sh}(W_1,W_2) }
\left\{\mu I_p(X;Y|U)+I_p(U;Z)\right\},
\\
&&{\cal C}_{\rm sh}(W_1,W_2)
\\
&=& 
\bigcap_{\mu>0}\{(R_1,R_2):\mu R_1+R_2 
\leq C^{(\mu)}(W_1,W_2)\}.
\eeqno 
Then we have the following
\beqno
{\cal C}(W_1,W_2)={\cal C}_{\rm sh}(W_1,W_2).
\eeqno
\end{itemize}
\end{pr}

Property \ref{pr:pro0} is a well known result.
We omit the proof of this property.
}
%
%
\newcommand{\ApdaAAA}{
\subsection{
Cardinality Bound on Auxiliary Random Variables
}
\label{sub:ApdaAAA}
}
\newcommand{\ApdaAAAzz}{ 

We first prove the following lemma.
\begin{lm}
\label{lm:CardLm}
\beqno
& &\overline{C}^{(\mu)}(W_1,W_2)
\\
&\defeq &\max_{\scs p \in {\cal P}(W_1,W_2)}
\left\{\mu I_p(X;Y|U) + I_p(U;Z) \right\}
\\
&=&C^{(\mu)}(W_1,W_2)
\\
&\defeq &\max_{\scs p \in {\cal P}(W_1,W_2)}
\left\{\mu I_p(X;Y|U) + I_p(U;Z) \right\}.
\eeqno
\end{lm}

{\it Proof:} We bound the cardinality $|{\cal U}|$ of $U$ 
to show that the bound $|{\cal U}|\leq |{\cal X}|$ is 
sufficient to describe $\overline{C}^{(\mu)}(W_1,W_2)$. 
Observe that 
\beqa
\hspace*{-5mm}& &p_{X}(x)
=\sum_{u\in {\cal U}}p_U(u)
p_{{X}|{U}}(x|u),
\label{eqn:asdfq}
\\
\hspace*{-5mm}
& &\mu I_p(X;Y|U)+I_p(U;Z)=\sum_{u\in {\cal U}}p_{U}(u)
\pi(p_{X|U}(\cdot|u)),
\label{eqn:aqqqaq}
\eeqa
where we set
\beqno
&&\pi(p_{{X}|U}(\cdot|u))
\defeq \sum_{(x,y)\in{\cal X}\times{\cal Y}}
p_{X|U}(x|u) W_1(y|x)W_2(z|y)
\\
& &\times \log \left\{
\frac{
\ds \frac{\ds 
\sum_{\tilde{x} \in {\cal X}}
\sum_{\tilde{y} \in {\cal Y}}W_2(z|\tilde{y})
W_1(\tilde{y}|\tilde{x}) p_{X|U}(\tilde{x}|u)}
{\ds
\sum_{\tilde{x} \in {\cal X}}
\sum_{\tilde{y} \in {\cal Y}}W_2(z|\tilde{y})
W_1(\tilde{y}|\tilde{x}) p_{X}(\tilde{x})
}
}
{\ds \frac{\ds \left[\sum_{\tilde{x} \in {\cal X}}
W_1(y|\tilde{x}) p_{X|U}(\tilde{x}|u)\right]^{\mu}}
{W_1^\mu(y|x)}
}
\right\}.
\eeqno
Here we remark that the quantities $p_{X}(\cdot)$ 
appearing in the above definition of $\pi(p_{X|U}(\cdot|u))$ 
is regarded as constants under (\ref{eqn:asdfq}). 
For each $u\in {\cal U}$, $\pi(p_{{X}|U}(\cdot|u))$ is a 
continuous function of $p_{X|U}(\cdot|u)$. Then by 
the support lemma,
$$
|{\cal U}| \leq |{\cal X}|-1 +1= |{\cal X}| 
$$
is sufficient to express $|{\cal X}|-1$ values 
of (\ref{eqn:asdfq}) and 
one value of (\ref{eqn:aqqqaq}). 
\hfill \IEEEQED
} 
\newcommand{\ApdaAAB}{
\subsection{Proof of Property \protect{\ref{pr:pro0}}}
\label{sub:ApdaAAB}


In this appendix we prove Property \ref{pr:pro0}. Property \ref{pr:pro0} 
part a) is a well known property. Proof of this property 
is omitted here. 

\begin{lm}\label{lm:asgsq} 
Suppose that  
$(\hat{R}_1,\hat{R}_2)$ 
does not belong to ${\cal C}(W_1,W_2)$. 
Then there exist $\epsilon, \mu^* >0$ such that 
for any $(R_1,R_2)\in {\cal C}(W_1,W_2)$
we have  
\beqno
&&\:\:\, \mu^*(R_1-\hat{R}_1)
  +(R_2-\hat{R}_2)
+\epsilon \leq 0.
\eeqno
\end{lm} 

Proof of this lemma is omitted here. Lemma \ref{lm:asgsq} 
is equivalent to the fact that if the region ${\cal C}(W_1,W_2)$ is a convex 
set, then for any point $(\hat{R}_{1},\hat{R}_2)$ outside ${\cal C}(W_1,W_2)$, 
there exits a line which
separates the point $(\hat{R}_{1},\hat{R}_2)$ from 
the region ${\cal C}(W_1,W_2)$. 
%
%

Proof of this lemma is omitted here. This lemma will be used 
to prove Property \ref{pr:pro0} part b). 

{\it Proof of Property \ref{pr:pro0} part b):} 
We first recall the following definitions of ${\cal P}(W_1,W_2)$ 
and ${\cal P}_{\rm sh}(W_1,W_2)$: 
\beqno
{\cal P}(W_1,W_2)
&\defeq& 
\{p_{UXY}: \pa {\cal U} \pa \leq 
\pa {\cal X}\pa+1,
U \markov  X\markov Y \},
\\
{\cal P}_{\rm sh}(W_1,W_2)
&\defeq& 
\{
p_{UXY}: \pa {\cal U} \pa \leq \pa{\cal X}\pa,
U \markov  X\markov Y \}.
\eeqno
We first prove 
${\cal C}_{sh}($ $W_1,W_2)$ 
$\subseteq $ ${\cal C}$ $(W_1,W_2)$. 
We assume that 
$(\hat{R}_1,$ $\hat{R}_2) \notin $ 
${\cal C}(W_1,W_2)$.
Then by Lemma \ref{lm:asgsq}, 
there exist $\epsilon, \mu^*>0$ such that 
for any $(R_1,R_2) \in {\cal C}_{\rm ext}(W_1,W_2)$
we have  
\beqno
&&\:\:\, \mu^*(R_1-\hat{R}_1)+(R_2-\hat{R}_2)
+\epsilon \leq 0.
\eeqno
Then we have
\beqa
& &\mu^* \hat{R}_1+ \hat{R}_2
\nonumber\\
&\geq& \max_{\scs R_1,R_2)
   \atop{\scs \in {\cal C}(W_1,W_2)}} 
   \left\{\mu^* R_1+ R_2\right\}+\epsilon
\nonumber\\ 
&\MEq{a}& \max_{p\in {\cal P}(W_1,W_2)}
\{ \mu^* I_p(X;Y|U)+I_p(U;Z)\}+\epsilon
\nonumber\\
&\geq& \max_{p\in {\cal P}_{\rm sh}(W_1,W_2)}
\{\mu^* I_p(X;Y|U)+I_p(U;Z)\}+\epsilon
\nonumber\\
&=&C^{(\mu^*)}(W_1,W_2)+\epsilon.
\label{eqn:sddsd}
\eeqa
Step (a) follows from the definition of 
${\cal C}(W_1,W_2)$. 
The bound (\ref{eqn:sddsd}) implies that 
$(\hat{R}_1,\hat{R}_2)$ 
$\notin {\cal C}_{\rm sh}(W_1,W_2)$.
Thus ${\cal C}_{\rm sh}(W_1,W_2)
\subseteq {\cal C}(W_1,W_2)$ 
is proved. We next prove 
${\cal C}($ $W_1,W_2)$ $\subseteq $ 
${\cal C}_{\rm sh}$$(W_1,W_2)$. We assume that 
$(R_1,$ $R_2) \in$ ${\cal C}(W_1,W_2)$.
Then there exists $p\in$ ${\cal P}$ $(W_1,W_2)$ such that
\beq
R_1 \leq I_p(X;Y|U), R_2 \leq I_p(U;Z).
\label{eqn:assz0}
\eeq
Then, for $(R_1,R_2)$ $\in {\cal C}(W_1,W_2)$, 
we have the following chain of inequalities:
\beqno
& & \mu R_1+ R_2
\\
&\MLeq{a}&\mu I_p(X;Y|U) + I_p(U;Z) 
\\
&\leq&
\max_{p \in {\cal P}(W_1,W_2)}
\ba[t]{l}\{ \mu I_p(X;Y|U)+I_p(U;Z)\}
\ea
\vSpa\\
\\
&\MEq{b}&
\max_{p \in {\cal P}_{\rm sh}(W_1,W_2)}
\ba[t]{l}\{\mu I_p(X;Y|U)+I_p(U;Z)\}\ea
\vSpa\\
&=&{C}^{(\mu)}(W_1,W_2).
\eeqno
Step (a) follows from (\ref{eqn:assz0}).
Step (b) follows from that by Lemma \ref{lm:CardLm}
stating that the cardinality bound in 
${\cal P}_{\rm }(W_1,W_2)$ can be reduced to that in ${\cal P}_{\rm sh}(W_1,W_2)$. 
Hence we have ${\cal C}(W_1,W_2)\subseteq {\cal C}_{\rm sh}(W_1,W_2)$. 
\hfill\IEEEQED
}
%
%
The broadcast channel was posed and investigated by Cover $\cite{cov72}$.
Bergmans \cite{bgm73} proved that ${\cal C}(W_1,W_2)$ serves as an 
inner bound of ${\cal C}_{\rm DBC}(W_1,W_2)$. 
Gallager \cite{gal74}, Ahlswede and K\"orner \cite{ak75}, 
proved that the inner bound ${\cal C}(W_1,W_2)$ is tight, thereby 
establishing the following theorem.   

\begin{Th}[Gallager \cite{gal74},Ahlswede and K\"orner \cite{ak75}]
\label{th:ddirect}{\rm ${}$ \\ For any DBC $(W_1,W_2)$, we have
$$
{\cal C}_{\rm DBC}(W_1,W_2)={\cal C}(W_1,W_2).
$$
}
\end{Th}

The strong converse theorem was proved by 
Ahlswede {\it et al.} \cite{agk76}. 
Their result is the following:
\begin{Th}[Ahlswede { et al.} \cite{agk76}] For each fixed \\
$(\varepsilon_1,\varepsilon_2)$ $ \in (0,1)^2$ and any DBC 
$(W_1,W_2)$, we have 
$$
 {\cal C}_{\rm m, DBC}(\varepsilon_1,\varepsilon_2|W_1,W_2)
={\cal C}_{\rm DBC}(W_1,W_2).
$$
\end{Th}

Their method used to prove the strong converse theorem was 
extended to the method called the image size 
characterization by Csisz\'ar and K\"orner \cite{ck}.

To examine an asymptotic behavior of ${\rm P}_{\rm c}^{(n)}$ 
for rates outside the capacity region ${\cal C}(W_1,W_2)$,
we define the following quantity. 
\beqno
& & 
G^{(n)}(R_1,R_2|W_1,W_2)
\\
&&\defeq
\min_{\scs 
(\varphi^{(n)},\psi_1^{(n)},\psi_2^{(n)}):
    \atop{\scs 
         (1/n)\log | {\cal K}_n |\geq R_1,
         \atop{\scs 
         (1/n)\log | {\cal L}_n |\geq R_2
         }
    }
}
\hspace*{-2mm}
\left(-\frac{1}{n}\right)
\log {\rm P}_{\rm c}^{(n)}
(\varphi^{(n)},\psi_1^{(n)},\psi_2^{(n)}),
\\
& &G(R_{1},R_2|W_1,W_2) 
\defeq \lim_{n\to\infty}G^{(n)}(R_{1},R_2|W_1,W_2). 
\eeqno
\newcommand{\OmitZ}{
By time sharing we have that 
\beqno
& &G^{(n+m)}\left(\left.
\frac{n R_1+m R_1^{\prime}}{n+m},
\frac{n R_2+m R_2^{\prime}}{n+m}\right|W_1,W_2\right) 
\\
&\leq& \frac{nG^{(n)}(R_1,R_2|W_1,W_2) 
+mG^{(m)}(R_1^{\prime},R_2^{\prime}|W_1,W_2)}{n+m}
\label{eqn:aaZ}. 
\eeqno
Choosing $R=R^\prime$ in (\ref{eqn:aaZ})  
we 
obtain the following subadditivity property
on $\{G^{(n)}(R_1,R_2|W_1,W_2)$ $\}_{n\geq 1}$: 
\beqno
& &G^{(n+m)}(R_1,R_2|{W}_1,W_2) 
\\
&\leq& \frac{nG^{(n)}(R_1,R_2|W_1,W_2) 
+mG^{(m)}(R_1,R_2|W_1,W_2)}{n+m},
\eeqno
from which we have that $G(R_{1},R_2|W_1,W_2)$ exists and 
satisfies the following.  
\beqno
\lim_{n\to\infty}G^{(n)}(R_{1},R_2|W_1,W_2) 
=\inf_{n\geq 1}G^{(n)}(R_1,R_2|W_1,W_2).
\eeqno
The exponent function $G(R_1,R_2|W_1,W_2)$ is a convex function 
of $(R_1,R_2)$. In fact, from (\ref{eqn:aaZ}), we have that 
for any $\alpha \in [0,1]$
\beqno
& &G(\alpha R_1+\bar{\alpha}R_1^{\prime},
     \alpha R_2+\bar{\alpha}R_2^{\prime}|W_1,W_2)
\\
&\leq &
\alpha G(R_1, R_2|W_1,W_2)
+\bar{\alpha} G( R_1^{\prime},R_2^{\prime}|W_1,W_2).
\eeqno
The region ${\cal R}(W_1,W_2)$ is also a closed convex set. }
Our main aim is to find an explicit 
In this paper we derive an explicit lower bound of $G(R_1,R_2|(W_1,W_2)$ 
that is positive if and only if $(R_1,R_2)\notin {\cal C}(W_1,W_2)$. 

\section{Main Result}

In this section we state our main result. Define 
\beqno
& &\omega^{(\mu)}_{q}(x,y,z|u)
\\
&\defeq&\mu\log \frac{q_{Y|X}(y|x) }{q_{Y|U}(y|u)}
+\log \frac{q_{Z|U}(z|u)}{q_{Z}(z)},
\\
& &\Lambda_{q}^{(\mu,\lambda)}({XYZ|U})
\\
&\defeq &\sum_{(u,x,y,z)\in 
{\cal U} \times 
{\cal X} \times 
{\cal Y} \times 
{\cal Z}
}q_{UX}(u,x)q_{Y|X}(y|x)q_{Z|Y}(z|y)
\\
&&\quad \times
\exp\left\{\lambda \omega^{(\mu)}_{q}(x,y,z|u) \right\},
\\
& &\Omega^{(\mu,\lambda)}_{q}(XYZ|U)
\defeq\log \Lambda_{q}^{(\mu,\lambda)}({XYZ|U}),
\\
& &\Omega^{(\mu,\lambda)}(W_1,W_2)
\defeq \max_{q \in{\cal P}_{\rm sh}(W_1,W_2)}
\Omega_{q}^{(\mu,\lambda)}({XYZ|U}),
\\
& &
F^{(\mu,\lambda)}(\mu R_1+R_2|W_1,W_2)
\\
&\defeq&
\frac{\lambda(\mu R_1+R_2)-\Omega^{(\mu,\lambda)}(W_1,W_2)}
{1+2\lambda+\lambda\mu},
\\
& &F(R_1,R_2|W_1,W_2)
\\
&\defeq&\sup_{\mu,\lambda >0} 
F^{(\mu,\lambda)}(\mu R_1+R_2|W_1,W_2).
\eeqno
%
%
%
%
We can show that the above functions and sets 
satisfy the following property. 
\begin{pr}\label{pr:pro1}  
$\quad$
\begin{itemize}
\item[a)]
For each $q \in {\cal P}(W_1,W_2)$, 
$\Omega_{q}^{(\mu,\lambda)}(XYZ|U)$ is a monotone increasing and 
convex function of $\lambda>0$. 
\item[b)] For every $q \in {\cal P}_{\rm sh}(W_1,W_2)$, we have 
\beqno
& &\lim_{\lambda\to +0}
\frac{\Omega_{q}^{(\mu,\lambda)}(XYZ|U)}{\lambda}
\\
&&=\mu I_q(X;Y|U)+I_q(U;Z).
\eeqno
\item[c)] If $(R_1,R_2) \notin {\cal C}(W_1,W_2)$,
then we have $F(R_1,$ $R_2|$ $W_1,W_2)>0$. 
%
%
%
%
%
\end{itemize}

\end{pr}

Proof of Property \ref{pr:pro1} is given in Appendix \ref{sub:ApdaAAC}. 
\newcommand{\ApdaAAC}{
\subsection{
Proof of Property \ref{pr:pro1}
} 
\label{sub:ApdaAAC}

In this appendix we prove Property \ref{pr:pro1}. 

{\it Proof of Property \ref{pr:pro1}:} 
We first prove part a) and b). 
For simplicity of notations, set
\beqno
& & 
\underline{a} \defeq (u,x,y,z), \underline{A}\defeq (U,X,Y,Z),
\underline{\cal A} \defeq 
 {\cal U} \times 
 {\cal X} \times 
 {\cal Y} \times 
 {\cal Z},
\\
& & \omega^{(\mu)}_{q}(x,y,z|u) \defeq \rho(\underline{a}),
\Omega_{q}^{(\mu,\lambda)}({XYZ|U})\defeq \xi(\lambda).
\eeqno
Then we have 
$$
\Omega^{(\mu,\lambda)}_{q}(XYZ|U)=\xi(\lambda)=\log
\left
[\sum_{\underline{a}\in \underline{\cal A} }q_{\underline{A}}(\underline{a})
{\rm e}^{\lambda\rho(\underline{a})}\right].
$$
By simple computations we have 
\beqa
& &\xi^{\prime}(\lambda)=
{\rm e}^{-\xi(\lambda)}
\left[\sum_{\underline{a}} q_{\underline{A}}(\underline{a})
\rho(\underline{a})
{\rm e}^{\lambda\rho(\underline{a})}\right],
\label{eqn:aaz11}
\\
& & \xi^{\prime\prime}(\lambda)=
{\rm e}^{-2\xi(\lambda)}
\nonumber\\
&&\times 
\left[\sum_{\underline{a}, 
            \underline{b}\in \underline{\cal A}}
 q_{\underline{A}}(\underline{a})
 q_{\underline{A}}(\underline{b})
 \frac{\left\{\rho(\underline{a})
       -\rho(\underline{b})\right\}^2}{2}
{\rm e}^{\lambda\left\{\rho(\underline{a})
         +\rho(\underline{b})\right\}}\right].\quad 
\label{eqn:aaz011}
\eeqa
From (\ref{eqn:aaz011}), it is obvious that 
$\xi^{\prime\prime}(\lambda)$ is nonnegative. Hence 
$\Omega_{q}^{(\mu,\lambda)}(XYZ|U)$ 
is a convex function of $\lambda$. 
It follows from (\ref{eqn:aaz11}) that for each 
$q \in {\cal P}(W_1,W_2)$, 
we have
\beqa
\xi^{\prime}(0)
&=&\sum_{\underline{a}} q_{\underline{A}}(\underline{a})
   \rho(\underline{a})
\nonumber\\
&=&\mu I_q(X;Y|U)+I_q(U;Z)\geq 0. 
\label{eqn:sddd}
\eeqa
Hence we have the part b). Since $\xi^{\prime}(0)\geq 0$ and 
$\xi^{\prime\prime}(\lambda)\geq 0$, we have 
$\xi^{\prime}(\lambda)\geq 0$ for $\lambda>0$.  
Hence for each $q \in {\cal P}(W_1,W_2)$, 
$\Omega_{q}^{(\mu,\lambda)}(XYZ|U)$ is monotone 
increasing for $\lambda>0$. Next we prove the part c).
We assume that $(R_1,R_2) \notin {\cal C}(W_1,W_2)$, 
then by Property \ref{pr:pro0} part b), 
there exist $\mu^* >0 $ and $\epsilon >0$, 
such that 
\beq
{\mu^* R_1+R_2 \geq C^{(\mu^*)}(W_1,W_2)} + \epsilon.
\label{eqn:ssdb}
\eeq
Set 
$$
\zeta(\lambda) \defeq 
\xi(\lambda)-\lambda\left[
I_q(X;Y|U)+I_q(U;Z)+\frac{\epsilon}{2}
\right].
$$
Then we have the following:
\beq
\zeta(0)=0, 
\ds \zeta^{\prime}(0)=-\frac{\epsilon}{2},
\zeta^{\prime\prime}(\lambda)=\xi^{\prime\prime}(\lambda)\geq 0.
\label{eqn:aaasd}
\eeq
It follows from (\ref{eqn:aaasd}) that 
there exists $\nu(\epsilon)>0$ 
such that we have $\zeta(\lambda)\leq 0$ 
for $\lambda\in (0, \nu(\epsilon)]$. 
Hence for any $\lambda\in (0, \nu(\epsilon)]$, 
for any $\mu \geq 0$, and for every 
$q \in {\cal P}_{\rm sh}(W_1,W_2)$, we have 
\beqa
\Omega^{(\mu,\lambda)}_{q}(UXYZ)
\leq \lambda \left(\mu I_q(X;Y|U)+I_q(U;Z)+\frac{\epsilon}{2}\right). 
\label{eqn:zzz012}
\eeqa
From (\ref{eqn:zzz012}), we have that for any 
$\lambda\in (0,\nu(\epsilon)]$ and for any $\mu>0$, 
\beqa
& &\Omega^{(\mu,\lambda)}(W_1,W_2)
\nonumber\\
&=&\max_{q \in {\cal P}_{\rm sh}(W_1,W_2)}
   \Omega^{(\mu,\lambda)}_{q}(UXYZ)
\nonumber\\
&\leq &\lambda
\left[
\max_{q\in {\cal P}_{\rm sh}(W_1,W_2)} 
\left\{\mu I_q(X;Y|U)+I_q(U;Z)
\right\}
+\frac{\epsilon}{2}
\right]
\nonumber\\
&= & \lambda \left[ 
\max_{q \in {\cal P}_{\rm sh}(W_1,W_2) } 
\left\{
\mu I_q(X;Y|U)+I_q(U;Z)\right\}
+\frac{\epsilon}{2}
\right]
\nonumber\\
&= & \lambda\left[ 
C^{(\mu)}(W_1,W_2)+\frac{\epsilon}{2}\right].
\label{eqn:ssda}
\eeqa
Under (\ref{eqn:ssdb}) and (\ref{eqn:ssda}), 
we have the following chain of inequalities: 
\beqno
& & F(R_1, R_2|W_1,W_2)
\\
&=&\sup_{\lambda>0}\sup_{\mu>0}
F^{(\mu,\lambda)}(\mu R_1+R_2|W_1,W_2)
\\
&\geq &\sup_{\lambda\in (0, \nu(\epsilon)]}
F^{(\mu^{*},\lambda)}(\mu^{*} R_1+R_2|W_1,W_2)
\\
&=&
\sup_{\lambda\in (0, \nu(\epsilon)]}
\frac{\lambda(\mu^* R_1+R_2)
-\Omega^{(\mu^*,\lambda)} (W_1,W_2)}
{1+2\lambda+\lambda \mu^*}
\\
&\MGeq{a} & 
\sup_{\lambda\in (0, \nu(\epsilon)]} 
\lambda
\frac{\ds \mu^* R_1+R_2-C^{(\mu^*)}(W_1,W_2)-\frac{\epsilon}{2}}
{1+2\lambda+\lambda \mu^*}
\\
&\MGeq{b}& 
\sup_{\lambda\in (0, \nu(\epsilon)]}
\frac{1}{2}\cdot \frac{\lambda \epsilon }
{1+2\lambda+\lambda \mu^*}
\\
&=&
\frac{1}{2}\cdot\frac{\nu(\epsilon)\epsilon }
{1+2\nu(\epsilon)+ \nu(\epsilon) \mu^*}>0.
\eeqno
Step (a) follows from (\ref{eqn:ssda}).
Step (b) follows from (\ref{eqn:ssdb}).
\hfill\IEEEQED
}
Our main result is the following. 
\begin{Th}\label{Th:main} 
For any degraded BC $(W_1,W_2)$, we have 
\beqa
G(R_1,R_2|W_1,W_2) &\geq& F(R_1,R_2|W_1,W_2).
\label{eqn:mainIeq}
\eeqa
\end{Th}

Proof of this theorem will be given in Section \ref{sec:Secaa}.
It follows from Theorem \ref{Th:main} and 
Property \ref{pr:pro1} part c) 
that if $(R_1,R_2)$ is outside the capacity region, 
then the error probability of decoding goes to one 
exponentially and its exponent is not below 
$F(R_1,R_2|W_1,W_2)$.
From this theorem we immediately obtain the following
corollary, which partially recovers the strong 
converse theorem by Ahlswede {\it et al.} \cite{agk76}. 
\begin{co} For each pair $(\varepsilon_1,\varepsilon_2)$ $ \in (0,1)^2$ 
satisfying $\varepsilon_1+\varepsilon_2$ $<1$, we have 
\beqno
& &{\cal C}_{\rm m, DBC}(\varepsilon_1,\varepsilon_2|W_1,W_2)
\\
&=&{\cal C}_{\rm DBC}(\varepsilon_1,\varepsilon_2|W_1,W_2)
  ={\cal C}_{\rm DBC}(\varepsilon_1+\varepsilon_2|W_1,W_2)
\\
&=&{\cal C}_{\rm DBC}(W_1,W_2)={\cal C}(W_1,W_2).
\eeqno
In particular, for each $\varepsilon \in (0,1/2)$, we have 
\beqno
&& {\cal C}_{\rm DBC}(\varepsilon,\varepsilon|W)
  ={\cal C}_{\rm DBC}(2\varepsilon|W_1,W_2)
={\cal C}(W_1,W_2).
\eeqno
\end{co}

The exponent function at rates outside 
the channel capacity was derived by 
Arimoto \cite{ari} and Dueck and K\"orner \cite{dk}. The 
techniques used by them are not useful to prove Theorem \ref{Th:main}.
Some novel techniques based on 
the information spectrum method introduced by Han \cite{han} 
are necessary to prove this theorem.

\section{Proof of the Results}
\label{sec:Secaa}

We first prove the following lemma. 
\begin{lm}\label{lm:Ohzzz}
For any $\eta>0$ and for any $(\varphi^{(n)},\psi_1^{(n)},\psi_2^{(n)})$  
satisfying  
$
(1/n)\log |{\cal K}_n| \geq R_1,
$ 
$
(1/n)\log |{\cal L}_n| \geq R_2.
$
we have
\beqa
& &{\rm P}_{\rm c}^{(n)}
(\varphi^{(n)},\psi_1^{(n)},\psi_2^{(n)})
\leq p_{L_nX^nY^nZ^n}\hugel
\nonumber\\
& &
R_1\leq \frac{1}{n}\log
\frac{W_1^n(Y^n|X^n)W_2^n(Z^n|Y^n)}{q_{Y^nZ^n|L_n}(Y^n,Z^n|L_n)}+\eta
\label{eqn:asppb}\\
& &
R_2 \leq \left.
\frac{1}{n}\log\frac{p_{Z^n|L_n}(Z^n|L_n)}{\tilde{q}_{Z^n}(Z^n)}+\eta
\right\}
+2{\rm e}^{-n\eta}. 
\label{eqn:azsad}
\eeqa
In (\ref{eqn:asppb}), we can choose any 
conditional distribution $q_{Y^nZ^n|L_n}$ on 
${\cal Y}^n\times {\cal Z}^n$ 
given $L_n$ $\in {\cal L}_n$. 
In (\ref{eqn:azsad}) we can choose any probability 
distribution $\tilde{q}_{Z^n}$ on ${\cal Z}^n$. 
\end{lm}

Proof of this lemma is given in Appendix \ref{sub:Apda}.
\newcommand{\Apda}{
\subsection{
Proof of Lemma \ref{lm:Ohzzz}
}\label{sub:Apda}

In this appendix we prove Lemma \ref{lm:Ohzzz}.

{\it Proof of Lemma \ref{lm:Ohzzz}:} 
For $l\in {\cal L}_n$, set
\beqno
{\cal A}_1(l)& \defeq &
\{(x^n, y^n,z^n): 
\ba[t]{l}
W_2^n(z^n|y^n)W_1^n(y^n|x^n)
\\
\geq |{\cal K}_n| {\rm e}^{-n\eta}
q_{Y^nZ^n|L_n}(y^n,z^n|l)
\},
\ea
\\
{\cal A}_2(l)
&\defeq &
\{(x^n,y^n,z^n): 
\ba[t]{l}
p_{Z^n|L_n}(z^n|l)\geq |{\cal L}_n| {\rm e}^{-n\eta}
\tilde{q}_{Z^n}(z^n)\},
\ea
\\
{\cal A}(l)&\defeq&
{\cal A}_1(l)\cap {\cal A}_2(l). 
\eeqno
Then we have the following: 
\beqno
{\rm P}_{\rm c}^{(n)}
&=&\frac{1}{|{\cal K}_n| |{\cal L}_n|} 
\sum_{(k,l)\in {\cal K}_n \times {\cal L}_n }
\sum_{ {\scs (x^n,y^n,z^n)\in {\cal A}(l),
         \atop{
          \scs y^n   \in {\cal D}_1(k), 
          \scs z^n  \in  {\cal D}_2(l)
         }
     }
   }1  
\\
& &\times 
\varphi^{(n)}(x^n|k,l)W_1^n(y^n|x^n)W_2^n(z^n|y^n)
\\
&&+\frac{1}{|{\cal K}_n| |{\cal L}_n|} 
  \sum_{(k,l)\in {\cal K}_n \times {\cal L}_n }
  \sum_{\scs (x^n,y^n,z^n)\in {\cal A}^c(l):
       \atop{
       \scs y^n \in {\cal D}_1(k), 
            z^n \in {\cal D}_2(l)
       }
    }1  
\\
& &\times 
\varphi^{(n)}(x^n|k,l)W_1^n(y^n|x^n)W_2^n(z^n|y^n)
\\
&\leq& \sum_{i=0,1,2}\Delta_i,
\eeqno
where
\beqno
{\Delta}_0
&\defeq &\frac{1}{|{\cal K}_n| |{\cal L}_n|} 
\sum_{(k,l)\in {\cal K}_n \times {\cal L}_n }
\sum_{\scs (x^n,y^n,z^n)\in {\cal A}(l)}
\\
& &\times 
{p}_{X^nY^nZ^n|K_n,L_n}(x^n,y^n,z^n|k,l),
\\
{\Delta}_i
&\defeq &
  \frac{1}{|{\cal K}_n| |{\cal L}_n|} 
  \sum_{(k,l)\in {\cal K}_n \times {\cal L}_n }
  \sum_{\scs (x^n,y^n,z^n)\in {\cal A}_i^c(l),
       \atop{\scs 
            y^n \in {\cal D}_1(k), 
            z^n \in {\cal D}_2(l)
       }
    }1  
\\
& &\times 
{p}_{X^nY^nZ^n|K_n,L_n}(x^n,y^n,z^n|k,l)
\\
& &\mbox{ for }i=1,2.
\eeqno
By definition we have 
\beqa
& &{\Delta}_0
\nonumber\\
&=&{p}_{L_nX^nY^nZ^n}\hugel
\nonumber\\
& &
\frac{1}{n}\log |{\cal K}_n| \leq \frac{1}{n}\log
\frac{W_1^n(Y^n|X^n)}{q_{Y^n|L_n}(Y^n|L_n)}+\eta,
\nonumber\\
& &
\frac{1}{n}\log |{\cal L}_n| \leq \left.
\frac{1}{n}\log\frac{p_{Z^n|L_n}(Z^n|L_n)}{\tilde{q}_{Z^n}(Z^n)}+\eta
\right\}.
\label{eqn:azsadaba}
\eeqa
From (\ref{eqn:azsadaba}), it follows that 
if $(\varphi^{(n)},\psi_1^{(n)},\psi_2^{(n)})$  
satisfies   
$
(1/n)$ $\log |{\cal K}_n| \geq R_1,
$
$
(1/n)\log |{\cal L}_n| \geq R_2,
$
then the quantity $\tilde{\Delta}_0$ is upper bounded by 
the first term in the right members of (\ref{eqn:azsad}) 
in Lemma \ref{lm:Ohzzz}.
Hence it suffices to show 
$\tilde{\Delta}_i\leq {\rm e}^{-n\eta},i=1,2$ 
to prove Lemma \ref{lm:Ohzzz}. 
We first prove $\tilde{\Delta}_1\leq {\rm e}^{-n\eta}$. 
We have the following chain of inequalities: 
\beqno
{\Delta}_1
&= &
  \frac{1}{|{\cal K}_n| |{\cal L}_n|} 
  \sum_{(k,l)\in {\cal K}_n \times {\cal L}_n }
  \sum_{\scs 
        (x^n,y^n,z^n):
         \atop{\scs 
             y^n \in {\cal D}_1(k), 
             z^n \in {\cal D}_2(l)
                \atop{\scs 
                    W_1^n(y^n|x^n)W_2(z^n|y^n)
                    \atop{\scs   
                     <{\rm e}^{-n\eta}|{\cal K}_n| 
                         \atop{\scs 
                         \times q_{Y^nZ^n|L_n}(y^n,z^n|l)
                         }
                     }
                }
         }
    }1
\\
& &\times 
\varphi^{(n)}(x^n|k,l)
W_1^n(y^n|x^n)W_2^n(z^n|y^n) 
\\
&\leq &
  \frac{ {\rm e}^{-n\eta} }{|{\cal L}_n|} 
  \sum_{(k,l)\in {\cal K}_n \times {\cal L}_n }
  \sum_{\scs
         (x^n,y^n,z^n):
           \atop {\scs 
            y^n \in {\cal D}_1(k), 
            z^n \in {\cal D}_2(l)
       }
    }1 
\\
& &\times 
{\varphi}^{(n)}(x^n|k,l)
q_{Y^nZ^n|L_n}(y^n,z^n|l) 
\\
&=&
  \frac{ {\rm e}^{-n\eta} }{|{\cal L}_n|} 
  \sum_{(k,l)\in {\cal K}_n \times {\cal L}_n }
q_{Y^nZ^n|L_n}\left(
\left.{\cal D}_1(k)\times {\cal D}_2(l)
\right| l \right)  
\\
&\leq&
  \frac{ {\rm e}^{-n\eta} }{|{\cal L}_n|} 
  \sum_{ l \in {\cal L}_n }
  \sum_{ k\in {\cal K}_n }
  q_{Y^n|L_n}\left(\left.{\cal D}_1(k)\right| l \right) 
\\
&=&
  \frac{ {\rm e}^{-n\eta} }{|{\cal L}_n|} 
  \sum_{ l\in {\cal L}_n }
  q_{Y^n|L_n}
\left(\left. \bigcup_{k\in {\cal K}_n}{\cal D}_1(k)\right|l \right) 
\\
&\leq&
  \frac{ {\rm e}^{-n\eta} }{|{\cal L}_n|} 
  \sum_{ l\in {\cal L}_n }1= {\rm e}^{-n\eta}.
\eeqno
Next we prove ${\Delta}_2\leq {\rm e}^{-n\eta}$.
We have the following chain of inequalities:
\beqno
{\Delta}_2
&= &
  \frac{1}{|{\cal L}_n|} 
  \sum_{(k,l) \in {\cal K}_n \times {\cal L}_n }
  \sum_{\scs 
        (x^n,y^n,z^n):
         \atop{\scs 
             y^n \in {\cal D}_1(k), 
             z^n \in {\cal D}_2(l)
                \atop{\scs 
                    p_{Z^n|L_n}(z^n|l) <{\rm e}^{-n\eta}
                    \atop{\scs   
                      \times |{\cal L}_n| \tilde{q}_{Z^n}(z^n)
                    }
                }
         }
    }1
\\
& &\times 
{p}_{K_nX^nY^nZ^n|L_n}(k,x^n, y^n, z^n|l) 
\\
&\leq & 
  \frac{1}{|{\cal L}_n|} \sum_{l\in {\cal L}_n} 
  \sum_{\scs 
        z^n \in {\cal D}_2(l),
                \atop{\scs 
                    p_{Z^n|L_n}(z^n|l) <{\rm e}^{-n\eta}
                    \atop{\scs   
                      \times |{\cal L}_n| \tilde{q}_{Z^n}(z^n)
                    }
                }
    }
\sum_{k\in {\cal K}_n} 
\sum_{(x^n,y^n)\in {\cal X}^n\times {\cal Y}^n }1
\\
& &\times 
{p}_{K_nX^nY^nZ^n|L_n}(k,x^n, y^n, z^n|l) 
\\
&\leq & 
  \frac{1}{|{\cal L}_n|} 
  \sum_{l\in {\cal L}_n} 
  \sum_{\scs 
            z^n \in {\cal D}_2(l),
            \atop{\scs 
                p_{Z^n|L_n}(z^n|l) <{\rm e}^{-n\eta}
                   \atop{\scs   
                        \times |{\cal L}_n| \tilde{q}_{Z^n}(z^n)
                    }
                }
        }  
    p_{Z^n|L_n}(z^n|l) 
\\
&\leq & 
  {\rm e}^{-n\eta} 
   \sum_{l\in {\cal L}_n }\sum_{z^n \in {\cal D}_2(l)}\tilde{q}_{Z^n}(z^n)
\\
&=& {\rm e}^{-n\eta} 
   \sum_{l\in {\cal L}_n }\tilde{q}_{Z^n}\left({\cal D}_2(l)\right)
\\
&=&{\rm e}^{-n\eta}\tilde{q}_{Z^n}
\left(\bigcup_{l \in {\cal L}_n} {\cal D}_2(l) \right)\leq {\rm e}^{-n\eta}. 
\eeqno
Thus Lemma \ref{lm:Ohzzz} is proved 
\hfill\IEEEQED
%
}

For $t=1,2,$ $\cdots,n$, set 
\beqno
& & {\cal U}_t\defeq {\cal L}_n 
\times {\cal Y}^{t-1} 
\times {\cal Z}^{t-1}, 
{\cal V}_t\defeq {\cal L}_n \times {\cal Z}^{t-1},
\\
& &
{U}_t \defeq (L_n,Y^{t-1},Z^{t-1}) \in {\cal U}_t,
{V}_t \defeq (L_n,Z^{t-1}) \in {\cal V}_t,
\\
& & {u}_t \defeq (l,y^{t-1},z^{t-1})\in {\cal U}_t, 
    {v}_t \defeq (l,z^{t-1})\in {\cal V}_t. 
\eeqno
For each $t=1,2\cdots,l$, let $\kappa_t$ 
be a natural projection from 
${\cal U}_t$ onto ${\cal V}_t$.  
Using $\kappa_t$, we have $V_t=
\kappa_t(U_t),$ $t=1,2,\cdots,n$.
For each $t=1,2,\cdots,n$, let 
$
{\cal Q}(\empty{\cal U}_t $ 
$\times {\cal X} \times $
${\cal Y} \times {\cal Z})
$
be a set of all probability distributions on
$$ 
\empty{\cal U}_t \times {\cal X} \times 
   {\cal Y} \times {\cal Z}
={\cal L}_n \times {\cal X} \times {\cal Y}^t \times {\cal Z}^t.
$$
For $t=1,2,\cdots, n$, we simply write  
${\cal Q}_t$$=$${\cal Q}(\empty{\cal U}_t $ 
$\times {\cal X} \times {\cal Y} \times {\cal Z})$.
Similarly, for $t=1,2,\cdots, n$, we simply write 
${q}_t=$ ${q}_{U_tX_tY_tZ_t}$ 
$\in {\cal Q}_t$.
Set
\beqno
{\cal Q}^n&\defeq& 
\prod_{t=1}^n {\cal Q}_t=\prod_{t=1}^n {\cal Q}(
\empty{\cal U}_t\times {\cal X}\times{\cal Y}\times {\cal Z}),
\\
q^n & \defeq & \left\{ {q}_t \right\}_{t=1}^n 
\in {\cal Q}^n.
\eeqno
By Lemma \ref{lm:Ohzzz} and some computations 
we have the following lemma.

\begin{lm}\label{lm:Ohzzzb}
For any $\eta>0$, for any 
$(\varphi^{(n)},\psi_1^{(n)},\psi_2^{(n)})$  
satisfying 
$
(1/n)\log |{\cal K}_n| \geq R_1,
$ 
$
(1/n)\log |{\cal L}_n| \geq R_2,
$
and for any ${q}^n \in {\cal Q}^n$, 
we have
\beqa
& &{\rm P}_{\rm c}^{(n)}
({\varphi}^{(n)},\psi_1^{(n)},\psi_2^{(n)})
\leq p_{L_nX^nY^nZ^n}
\hugel
\nonumber\\
& &{
R_1\leq \frac{1}{n}
\sum_{t=1}^n
\log \frac{W_1(Y_t|X_t)}{{q}_{Y_t|\empty{U}_t}(Y_t|\empty{U}_t)}
+\eta,}
\nonumber\\
& &
R_2\leq
\frac{1}{n}
\sum_{t=1}^n\log
\frac{{p}_{Z_t|
\empty{V}_t}(Z_t|\empty{V}_t)}{q_{Z_t}(Z_t)}
+\eta \huger +2{\rm e}^{-n\eta},
\label{eqn:saQQa}
\eeqa
where for each $t=1,2,\cdots,n$, the conditional probability 
distribution ${q}_{Y_t|U_t}$ and the probability 
distribution ${q}_{Z_t}$ appearing in the first term 
in the right members of (\ref{eqn:saQQa}) are chosen so that 
they are induced by the joint distribution 
$q_t={q}_{U_tX_tY_tZ_t}$
$\in {\cal Q}_t$. 
\end{lm}

Proof of this lemma is given in 
Appendix \ref{sub:LemmaOhzzzb}.
\newcommand{\ApdLemmaA}{
\subsection{Proof of Lemma \ref{lm:Ohzzzb}}
\label{sub:LemmaOhzzzb}

From Lemma \ref{lm:Ohzzz}, we have the following lemma 
\begin{lm}\label{lm:OhzzzB}
For any $\eta>0$ and for any 
$(\varphi^{(n)},\psi_1^{(n)},\psi_2^{(n)})$  
satisfying  
$
(1/n)\log |{\cal K}_n| \geq R_1,
$
$
(1/n)\log |{\cal L}_n| \geq R_2,
$
we have
\beqa
& &{\rm P}_{\rm c}^{(n)}
(\varphi^{(n)},\psi_1^{(n)},\psi_2^{(n)})
 \leq {p}_{L_n X^nY^nZ^n}
\hugel
\nonumber\\
& &{
R_1\leq \frac{1}{n}
\sum_{t=1}^n
\log \frac{W_1(Y_t|X_t)}{q_{Y_t|L_nY^{t-1}}(Y_t|L_n,Y^{t-1},Z^{t-1})}+\eta,}
\nonumber\\
& &
R_2\leq
\frac{1}{n}
\sum_{t=1}^n\log
\frac{{p}_{Z_t|{L_n}Z^{t-1}}
(Z_t|L_n,Z^{t-1})}{\tilde{q}_{Z_t}(Z_t)}
+\eta 
\huger +2{\rm e}^{-n\eta}.
\nonumber
\eeqa
\end{lm}

{\it Proof:} 
In (\ref{eqn:asppb}) in Lemma \ref{lm:Ohzzz}, 
we choose $q_{Z^nY^n|L_n}$ 
\beqno
& &q_{Y^nZ^n|L_n}(y^n,z^n|l)
\\
&=&\prod_{t=1}^n 
\left\{
q_{Y_t|L_n Y^{t-1} Z^{t-1}}(y_t|l,y^{t-1},z^{t-1}) \right.
\\ 
& & \qquad \times \left. q_{ Z_t| L_n Y^t Z^{t-1} }(z_t|l,y^t,z^{t-1}) \right\}
\\
&=&\prod_{t=1}^n \{q_{Y_t|L_n Y^{t-1} Z^{t-1}}
(y_t|l,y^{t-1},z^{t-1})W_2(z_t|y_t)\}.
\eeqno
In (\ref{eqn:azsad}) in Lemma \ref{lm:Ohzzz}, 
we choose $\tilde{q}_{Z^n}$ having the form 
$$
\tilde{q}_{Z^n}(Z^n)=\prod_{t=1}^n\tilde{q}_{Z_t}(Z_t).
$$
Then from the bound (\ref{eqn:azsad}) 
in Lemma \ref{lm:Ohzzz}, we obtain 
\beqa
& &{\rm P}_{\rm c}^{(n)}
(\varphi^{(n)},\psi_1^{(n)},\psi_2^{(n)})
\leq {p}_{L_nX^nY^nZ^n}\hugel
\nonumber\\
& &
R_1\leq \frac{1}{n}
\sum_{t=1}^n
\log \frac{W_1(Y_t|X_t)}{q_{Y_t|L_nY^{t-1}Z^{t-1}}
(Y_t|L_n,Y^{t-1},Z^{t-1})}+\eta,
\nonumber\\
& &
R_2\leq
\frac{1}{n}
\sum_{t=1}^n\log
\frac{p_{Z_t|{L_n}Z^{t-1}}(Z_t|L_n,Z^{t-1})}
{\tilde{q}_{Z_t}(Z_t)}
+\eta \huger +2{\rm e}^{-n\eta},
\nonumber
\eeqa
completing the proof. \hfill\IEEEQED

From Lemma \ref{lm:OhzzzB}, we immediately obtain 
Lemma \ref{lm:Ohzzzb}.
}
%


\newcommand{\ApdLemmaB}{
}
%

To evaluate an upper bound of (\ref{eqn:saQQa}) in Lemma \ref{lm:Ohzzzb}. 
We use the following lemma, which is well known as the Cram\`er's bound in 
the large deviation principle.
\begin{lm}
\label{lm:Ohzzzbz}
For any real valued random variable $Z$ and any $\theta>0$, 
we have
$$
\Pr\{Z \geq a \}\leq 
\exp
\left[
-\left(
\lambda a -\log {\rm E}[\exp(\theta Z)]
\right) 
\right].
$$
\end{lm}

Here we define a quantity which serves as an exponential
upper bound of ${\rm P}_{\rm c}^{(n)}(\varphi^{(n)},$ 
$\psi_1^{(n)},\psi_2^{(n)})$. 
Let ${\cal P}^{(n)}(W_1,W_2)$ be a 
set of all probability distributions 
${p}_{L_nX^nY^nZ^n}$ on 
${\cal L}_n$
$\times {\cal X}^n$
$\times {\cal Y}^n$
$\times {\cal Z}^n$
having the form:
\beqno
& &{p}_{L^nX^nY^nZ^n}(l,x^n,y^n,z^n)
\\
&=&{p}_{L^n}(l)
\prod_{t=1}^n 
{p}_{X_t|L_n X^{t-1}}
(x_t|l,x^{t-1})W_1(y_t|x_t)W_2(z_t|y_t).
\eeqno
For simplicity of notation we use the notation $p^{(n)}$ 
for $p_{L_nX^nY^nZ^n}$ $\in {\cal P}^{(n)}$
$(W_1,W_2)$. We assume that 
$
p_{U_tX_tY_tZ_t}=p_{L_nX_tY^{t}Z_{t}}
$
is a marginal distribution of $p^{(n)}$. 
For $t=1,2,\cdots, n$, we simply write $p_t=$ 
$p_{U_tX_tY_tZ_t}$. 
For $p^{(n)}$ $\in {\cal P}^{(n)}(W_1,W_2)$ 
and $q^n$ $\in {\cal Q}^n$, we define 
\beqno
&&
\Omega^{(\mu,\theta)}_{p^{(n)}||q^n}(X^nY^nZ^n|{L_n})
\\
&&
\defeq \log
{\rm E}_{p^{(n)}}
\left[
\prod_{t=1}^n
\frac{W_1^{\theta\mu}(Y_t|X_t)p^{\theta}_{Z_t|{V_t}}(Z_t|{V_t})}
{q^{\theta\mu}_{Y_t|{U_t}}(Y_t|{U_t})q^{\theta}_{Z_t}(Z_t)}
\right],
\eeqno
where for each $t=1,2,\cdots,n$, the conditional probability 
distribution $q_{Y_t|U_t}$ and the probability distribution 
$q_{Z_t}$ appearing in the definition of 
$\Omega^{(\mu,\theta)}_{p^{(n)}||q^{n}}$ 
$(X^nY^nZ^n|L_n)$ are chosen so that they 
are induced by the joint distribution 
$q_t=q_{U_tX_tY_tZ_t} \in {\cal Q}_t$.

Here we give a remark on an essential difference 
between $p^{(n)}$ $\in {\cal P}^{(n)}(W_1,W_2)$ 
and $q^n$ $\in{\cal Q}^n$. For the former the $n$ 
probability distributions $p_t,$ $t=1,2,\cdots, n,$ 
are consistent with $p^{(n)}$, since all of them are 
marginal distributions of $p^{(n)}$. On the other hand, 
for the latter, $q^{n}$ is just {\it a sequence} of 
$n$ probability distributions. Hence, we may not have 
the consistency between the $n$ elements $q_t$, 
$t=1,2,\cdots,n,$ of $q^n$. 

By Lemmas \ref{lm:Ohzzzb} and \ref{lm:Ohzzzbz}, we have 
the following proposition. 
\begin{pro}
\label{pro:Ohzzp}
For any $\mu,$ $\theta >0$, any $q^n \in {\cal Q}^n$, 
and any $(\varphi^{(n)},\psi_1^{(n)},\psi_2^{(n)})$  
satisfying  
\beq
\frac{1}{n}\log |{\cal K}_n| \geq R_1,
\frac{1}{n}\log |{\cal L}_n| \geq R_2,
\label{eqn:XasDD}
\eeq
we have 
\beqno	
& & {\rm P}_{\rm c}^{(n)}(\varphi^{(n)},\psi_1^{(n)},\psi_2^{(n)})
\\
&\leq &3\exp
\left\{
-n\frac{
\theta(\mu R_1+R_2)-\frac{1}{n}
{\Omega}_{p^{(n)}||q^{n}}^{(\mu,\theta)}(X^nY^nZ^n|{L_n})
}
{1+\theta+\theta\mu}
\right\}.
\eeqno
\end{pro}

{\it Proof:} Under the condition (\ref{eqn:XasDD}), 
we have the following chain of inequalities: 
\beqa
& &
{\rm P}_{\rm c}^{(n)}(\varphi^{(n)},\psi_1^{(n)},\psi_2^{(n)})
\MLeq{a} p_{{L_n}X^nY^nZ^n} \hugel
\nonumber\\
& &R_1\leq \frac{1}{n}
\sum_{t=1}^n
\log \frac{W_1(Y_t|X_t)}{q_{Y_t|{U_t}}(Y_t|U_t)}+\eta,
\nonumber\\
& & R_2 \leq 
\frac{1}{n}
\sum_{t=1}^n\log
\frac{p_{Z_t|{V_t}}(Z_t|{V_t})}{q_{Z_t}(Z_t)}+\eta
\huger
+3{\rm e}^{-n\eta} 
\nonumber\\ 
&\leq &p_{{L_n}X^nY^nZ^n}\hugel
\mu R_1+R_2-(\mu+1)\eta 
\nonumber\\
& &
\left. \leq \frac{1}{n}
\sum_{t=1}^n
\log \left[
\frac{W_1(Y_t|X_t) p_{Z_t|{V_t}}(Z_t|{V_t})}
{q^{\mu}_{Y_t|{U_t}}(Y_t|{U_t})
q^{\mu}_{Z_t}(Z_t)}
\right]\right\}
+3{\rm e}^{-n\eta} 
\nonumber\\
&\MLeq{b} &
\exp\Bigl[n\Bigl\{-\theta(\mu R_1+R_2)+\theta(\mu+1)\eta 
\nonumber\\
&&\qquad \left.\left.
+\frac{1}{n}\Omega_{p^{(n)}||q^{n}}^{(\mu,\theta)}
        (X^nY^nZ^n|{L_n})\right\}\right]
+3{\rm e}^{-n\eta}.
\label{eqn:aaabv}
\eeqa
Step (a) follows from Lemma \ref{lm:Ohzzzb}.
Step (b) follows from Lemma \ref{lm:Ohzzzbz}.
We choose $\eta$ so that 
\beqa
-\eta&=& -\theta(\mu R_1+R_2)+\theta(\mu+1)\eta 
\nonumber\\
& &   +\frac{1}{n}
\Omega_{p^{(n)}||q^{n}}^{(\mu,\theta)}(X^nY^nZ^n|{L_n}).
\label{eqn:aaappp}
\eeqa
Solving (\ref{eqn:aaappp}) with respect to $\eta$, we have 
\beqno
\eta=
\frac{
\theta(\mu R_1+R_2)-
\frac{1}{n}{\Omega}_{p^{(n)}||q^{n}}^{(\mu,\theta)}(X^nY^nZ^n|{L_n})
}
{1+\theta+\theta\mu}.
\eeqno
For this choice of $\eta$ and (\ref{eqn:aaabv}), we have
\beqno
& &{\rm P}_{\rm c}^{(n)}%
\leq 3{\rm e}^{-n\eta}
\\
&=&3\exp
\left\{
-n\frac{
\theta(\mu R_1+R_2)
-\frac{1}{n}{\Omega}_{p^{(n)}||q^{n}}^{(\mu,\theta)}(X^nY^nZ^n|{L_n})
}
{1+\theta+\theta\mu}
\right\},
\eeqno
completing the proof. 
\hfill \IEEEQED

Set 
\beqno
& &\overline{\Omega}^{(\mu,\theta)}(W_1,W_2)
\\
&\defeq & 
\sup_{n\geq 1}
\max_{\scs 
{p}^{(n)} \in {\cal P}^{(n)}(W_1,W_2)
}
\min_{\scs q^n \in {\cal Q}^n}
\frac{1}{n}\Omega_{p^{(n)}||q^{n}}^{(\mu,\theta)}(X^nY^nZ^n|{L_n}).
\eeqno
By the above definition of $G^{(n)}(R_1,$ $R_2|W_1,W_2)$ 
and Proposition \ref{pro:Ohzzp}, we have 
\beqa
& &G^{(n)}(R_1,R_2|W_1,W_2)
\nonumber\\
&\geq & 
\frac{
\theta(\mu R_1+R_2)-
\overline{\Omega}^{(\mu,\theta)}(W_1,W_2)
}{1+\theta+\theta\mu}
-\frac{1}{n}\log 3.
\label{eqn:aaaxc}
\eeqa
Then from (\ref{eqn:aaaxc}), we obtain the following corollary. 
\begin{co} 
\label{co:corOne}
For any $\theta >0,\mu>0$, we have 
$$
G(R_1,R_2|W_1,W_2)\geq 
\frac{\theta(\mu R_1+R_2)-\overline{\Omega}^{(\mu,\theta)}(W_1,W_2)
} {1+\theta+\theta\mu}.
$$
\end{co}

We shall call $\overline{\Omega}^{(\mu,\theta)}(W_1,W_2)$ 
the communication potential. The above corollary implies that 
the analysis of $\overline{\Omega}^{(\mu,\theta)}($ $W_1,W_2)$ 
leads to an establishment of a strong converse theorem 
for the degraded BC. 

\newcommand{\ApdLemmaC}{
\subsection{
Upper Bound of 
$\overline{\Omega}^{(\mu,\theta)}(W_1,W_2)$
} 

In this appendix we derive an explicit upper bound 
of $\overline{\Omega}^{(\mu,\theta)}(W_1,W_2)$ 
to prove Proposition \ref{pro:mainpro}.
For each $t=1,2,\cdots,n$, define the function of 
$(u_t,x_t,y_t,z_t)$
$\in {\cal U}_t$
$\times {\cal X}$
$\times {\cal Y}$
$\times {\cal Z}$ 
by 
\beqno
f_{p_t||q_t, \kappa_t}^{(\mu,\theta)}
(x_t,y_t,z_t|u_t)
&\defeq& 
\frac{W_1^{\theta\mu}(y_t|x_t)p_{Z_t|{U_t}}^{\theta}(z_t|u_t)}
{q_{Y_t|{U_t}}^{\theta \mu}(y_t|u_t) \tilde{q}_{Z_t}^{\theta}(z_t)}. 
\eeqno
For each $t=1,2,\cdots,n$, we define the probability distribution
\beqno
& &{p}_{L_nX^tY^tZ^t}^{(\mu,\theta;q^t, \kappa^t)}
\\
&\defeq& 
\left\{
{p}_{L_n X^tY^tZ^t}^{(\mu,\theta;q^t, \kappa^t)}(l,x^t,y^t,z^t)
\right\}_{(l,x^t,y^t,z^t)
\in {\cal L}_n \times {\cal X}^t \times {\cal Y}^t 
\times {\cal Z}^t}
\eeqno
by 
\beqno
& &
{p}_{L_n X^t Y^t Z^t}^{(\mu,\theta;q^t, \kappa^t)}(l,x^t,y^t,z^t) 
\\
&\defeq& C_t^{-1} p_{L_n}(l)p_{X^t|L_n}(x^t|l)
\prod_{i=1}^t \{W_1(y_i|x_i)W_2(z_i|y_i)
\\
& &\times f_{p_i||q_i, \kappa_i}^{(\mu,\theta)}(x_i,y_i,z_i|u_i)\},
\eeqno
where
\beqno
C_t&\defeq&
\sum_{l,x^t,y^t,z^t}
p_{L_n}(l)p_{X^t|L_n}(x^t|l)
\prod_{i=1}^t \{ W_1(y_i|x_i)W_2(z_i|y_i)
\\
& & \qquad \times 
     f_{p_i||q_i, \kappa_i}^{(\mu,\theta)}(x_i,y_i,z_i|u_i)\},
\eeqno
are constants for normalization. For each $t=1,2,\cdots,n$, set 
\beq 
\Phi_{t,q^t, \kappa^t}^{(\mu,\theta)}\defeq C_tC_{t-1}^{-1},
\label{eqn:defa}
\eeq
where we define $C_0=1$. Then we have the following lemma.
\begin{lm}\label{lm:keylm}
\beqa 
& &\Omega_{p^{(n)}||q^{n}}^{(\mu,\theta)}(X^nY^nZ^n|{L_n})
=\sum_{t=1}^n \log \Phi_{t,q^{t}}^{(\mu,\theta)}.
\label{eqn:defazzz}  
\eeqa
\end{lm}

{\it Proof}: From (\ref{eqn:defa}) we have
\beq
\log \Phi_{t,q^t, \kappa^t}^{(\mu,\theta)}
=\log C_t - \log C_{t-1}. 
\label{eqn:aaap}
\eeq
Furthermore, by definition we have 
\beq
\Omega^{(\mu,\theta)}_{p^{(n)}||q^{n}}(X^nY^nZ^n|{L_n})= \log C_n, C_0=1. 
\label{eqn:aaapq}
\eeq
From (\ref{eqn:aaap}) and (\ref{eqn:aaapq}), 
(\ref{eqn:defazzz}) is obvious. \hfill \IEEEQED

The following lemma is useful for the computation of 
$\Phi_{t,q^t, \kappa^t}^{(\mu,\theta)}$ for $t=1,2,\cdots,n$.
\begin{lm}\label{lm:aaa}
For each $t=1,2, \cdots,n$, and for any 
$(l,$ $x^t, y^t,z^t)\in {\cal L}_n$ 
$\times {\cal X}^t$
$\times {\cal Y}^t$
$\times {\cal Z}^t$,
we have
\beqa
& &p_{L_n X^t Y^t Z^t}^{(\mu,\theta;q^t, \kappa^t)}(l,x^t,y^t,z^t)
\nonumber\\
& &=(\Phi_t^{(\mu,\theta;q^t, \kappa^t)})^{-1}
p_{L_n X^{t-1} Y^{t-1} Z^{t-1} }^{(\mu,\theta;q^{t-1}, \kappa^{t-1})}
(l,x^{t-1},y^{t-1}, z^{t-1})
\nonumber\\
& &\quad \times 
p_{X_t|L_n,X^{t-1}}(x_t|l,x^{t-1})W_1(y_t|x_t)W_2(z_t|y_t)
\nonumber\\
& &\quad \times 
f_{p_t||q_t, \kappa_t}^{(\mu,\theta)}(x_t,y_t,z_t|u_t).
\label{eqn:satt}
\eeqa
Furthermore, we have
\beqa
& &\Phi_{t,q^t, \kappa^t}^{(\mu,\theta)}
\nonumber\\
&=& \sum_{l,x^t,y^t, z^t} 
p_{{L_n}X^{t-1}Y^{t-1}Z^{t-1}}
^{(\mu,\theta;q^{t-1}, \kappa^{t-1})}(l,x^{t-1},y^{t-1},z^{t-1})
\nonumber\\
& &\quad\times 
p_{X_t|L_n,X^{t-1}}(x_t|l,x^{t-1})W_1(y_t|x_t)W_2(z_t|y_t)
\nonumber\\
& &\quad\times 
f_{p_t||q_t, \kappa_t}^{(\mu,\theta)}
(x_t,y_t,z_t|u_t).
%
\label{eqn:sattb}
\eeqa
\end{lm}
}

\newcommand{\Apdc}{

{\it Proof of  Lemma \ref{lm:aaa}:} By the definition of 
${p}_{{L_n}X^tY^{t}Z^{t}}^{(\mu,\theta;q^t, \kappa^t)}$ $(l,$ $x^t,y^t,z^t)$, 
$t=1,2,\cdots,n$, we have 
\beqa
\hspace*{-4mm}
& &p_{{L_n}X^tY^{t}Z^{t}}^{(\mu,\theta;q^t, \kappa^t)}(l,x^t,y^{t},z^{t})
\nonumber\\
\hspace*{-4mm}
&=&C_t^{-1}
p_{L_n}(l)p_{X^t|L_n}(x^t|l)
\prod_{i=1}^t \{W_1(y_i|x_i)W_2(z_i|y_i)
\nonumber\\
& &
\times 
f_{p_i||q_i, \kappa_i}^{(\mu,\theta)}
(x_i,y_i,z_i|u_{i-1})\}.
\label{eqn:azaq}
\eeqa 
Then we have the following chain of equalities:
\beqa 
& &p_{{L_n}X^tY^tZ^t}^{(\mu,\theta;q^t, \kappa^t)}(l,x^t,y^t,z^t)
\nonumber\\
&\MEq{a}&
C_t^{-1} 
p_{L_n}(l)p_{X^t|L_n}(x^t|l)
\prod_{i=1}^t \{W_1(y_i|x_i)W_2(z_i|y_i)
\nonumber\\
& &
\times 
f_{p_i||q_i, \kappa_i}^{(\mu,\theta)}
(x_i,y_i,z_i|u_i)\}
\nonumber\\
&=&C_t^{-1}
p_{L_n}(l)p_{X^{t-1}|L_n}(x^{t-1}|l)
\prod_{i=1}^t \{W_1(y_i|x_i)W_2(z_i|y_i)
\nonumber\\
&&\times 
f_{p_i||q_i, \kappa_i}^{(\mu,\theta)}
(x_i,y_i,z_i|u_i)\}
\nonumber\\
& &\times 
p_{X_t|L_nX^{t-1}}(x_t|l,x^{t-1})
W_1(y_t|x_t)W_2(z_t|y_t) 
\nonumber\\
& &\times 
f_{p_t||q_t, \kappa_t}^{(\mu,\theta)}
(x_t,y_t, z_t|u_t)
\nonumber\\
&\MEq{b}&
C_t^{-1}C_{t-1}
p_{{L_n}X^{t-1}Y^{t-1}Z^{t-1}}^{(\mu,\theta;q^{t-1}, \kappa^{t-1})}
(l,x^{t-1},y^{t-1}, z^{t-1})
\nonumber\\
& &\times 
p_{X_t|L_nX^{t-1}}(x_t|l,x^{t-1})
W_1(y_t|x_t)W_2(z_t|y_t) 
\nonumber\\
& &\times 
f_{p_t||q_t, \kappa_t}^{(\mu,\theta)}
(x_t,y_t,z_t|u_t)
\nonumber\\
&=&(\Phi_{t,q^t, \kappa^t}^{(\mu,\theta)})^{-1}
p_{{L_n}X^{t-1}Y^{t-1}Z^{t-1}}^{(\mu,\theta;q^{t-1}, \kappa^{t-1})}
(l,x^{t-1},y^{t-1},z^{t-1})
\nonumber\\
& &\times 
p_{X_t|L_nX^{t-1}}(x_t|l,x^{t-1})
W_1(y_t|x_t)W_2(z_t|y_t) 
\nonumber\\
& &\times f_{p_t||q_t, \kappa_t}^{(\mu,\theta)}
(x_t,y_t,z_t|u_t).
\label{eqn:daaaq}
\eeqa
Steps (a) and (b) follow from (\ref{eqn:azaq}). 
From (\ref{eqn:daaaq}), we have 
\beqa 
& &\Phi_{t,q^t, \kappa^t}^{(\mu,\theta)}
p_{{L_n}X^tY^tZ^t}^{(\mu,\mu ;q^t, \kappa^t)}(l,x^t,y^t,z^t)
\label{eqn:daxx}\\
&=&p_{{L_n}X^{t-1}Y^{t-1}Z^{t-1}}^{(\mu,\theta;q^{t-1}, \kappa^{t-1})}
(l,x^{t-1},y^{t-1},z^{t-1})
\nonumber\\
& &\quad\times 
p_{X_t|L_nX^{t-1}}(x_t|l,x^{t-1})
W_1(y_t|x_t)W_2(z_t|y_t) 
\nonumber\\
& &\quad\times
f_{p_t||q_t, \kappa_t}^{(\mu,\theta)}
(x_t,y_t,z_t|u_t).
\label{eqn:daaxx}
\eeqa
Taking summations of (\ref{eqn:daxx}) and (\ref{eqn:daaxx}) 
with respect to $l,x^t,$ $y^t,$ $z^t$, we obtain 
\beqa
& &\Phi_{t,q^t, \kappa^t}^{(\mu,\theta)}
\nonumber\\
&=& \sum_{l,x^t,y^t,z^t}
p_{{L_n}X^{t-1}Y^{t-1}Z^{t-1}}^{(\mu,\theta;q^{t-1}, \kappa^{t-1})}
(l,x^{t-1},y^{t-1},z^{t-1})
\nonumber\\
& & \times p_{X_t|L_n,X^{t-1}}(x_t|l,x^{t-1})W_1(y_t|x_t)W_2(z_t|y_t)
\nonumber\\
& & \times 
f_{p_t||q_t, \kappa_t}^{(\mu,\theta)}(x_t,y_t,z_t|u_t),
\nonumber
\eeqa
completing the proof.
\hfill \IEEEQED
}
\newcommand{\ApdLemmaCContA}{

We set 
\beqno
& &p_{U_tX_t}^{(\mu,\theta;q^{t-1}, \kappa^{t-1})}(u_t,x_t)
=p_{L_nX_tY^{t-1}Z^{t-1}}^{(\mu,\theta;q^{t-1}, \kappa^{t-1})}(l,x_t,y^{t-1},z^{t-1})
\\
& &\defeq\sum_{x^{t-1}}
p_{{L_n}X^{t-1}Y^{t-1}Z^{t-1}}^{(\mu,\theta;q^{t-1}, \kappa^{t-1})}
(l,x^{t-1},y^{t-1},z^{t-1})
\\
& &\qquad \times p_{X_t|{L_n}X^{t-1}}(x_t|l,x^{t-1}).
\eeqno
Then by (\ref{eqn:sattb}) in Lemma \ref{lm:aaa} and the definition 
of
$f_{p_t||q_t, \kappa_t}^{(\mu,\theta)}$
$(x_t$$,y_t,$$z_t$
$|u_t)$, we have 
\beqa
& &
\Phi_{t,q^{t},\kappa^t}^{(\mu,\theta)}
\nonumber\\
&=&
\sum_{u_t,x_t, y_t,z_t}
p_{U_tX_t}^{(\mu,\theta;q^{t-1}, \kappa^{t-1})}(u_t,x_t)
W_1(y_t|x_t)W_2(z_t|y_t)
\nonumber\\
& &\quad \times 
\frac{W_1^{\theta\mu}(y_t|x_t)p_{Z_t|V_t}^{\theta}(z_t|v_t)}
{q_{Y_t|{U_t}}^{\theta \mu}(y_t|u_t){q}_{Z_t}^{\theta}(z_t)}. 
\label{eqn:aasor}
\eeqa
}
The following proposition is a mathematical core 
to prove our main result.
\begin{pro}\label{pro:mainpro}
For $\theta\in (0,1)$, set 
\beq
\lambda=\frac{\theta}{1-\theta} 
\Leftrightarrow \theta=\frac{\lambda}{1+\lambda}. 
\label{eqn:abaddd}
\eeq 
Then, for any $\theta \in (0,1)$, we have 
$$
\overline{\Omega}^{(\mu,\theta)}(W_1,W_2)
\leq \frac{1}{1+\lambda}\Omega^{(\mu,\lambda)}(W_1,W_2).
$$
\end{pro}

Proof of this proposition is in Appendix \ref{sub:mainPro}. 
The proof is not so simple. We must 
introduce a new method for the proof. 
\newcommand{\ApdLemmaCContB}{
\label{sub:mainPro}
Proof of Proposition \ref{pro:mainpro} is as follows. 

{\it Proof of Proposition \ref{pro:mainpro}:} 
Set
\beqno
&&\hat{\cal P}_n(W_1,W_2)
\defeq 
\{q:
\pa {\cal U} \pa \leq \pa {\cal L}_n\pa 
\pa {\cal Y}\pa^{n-1},
\vSpa\\
&&\qquad 
q_{Y|X}=W_1, q_{Z|Y}=W_2,  
U \markov X\markov Y \markov Z \},
\\
& &\hat{\Omega}_n^{(\mu,\lambda)}(W_1,W_2)
\defeq \max_{q \in \hat{\cal P}_n(W_1,W_2)}
\Omega_q^{(\mu,\lambda)}({XYZ|U}).
\eeqno
We choose $q_t=$ $q_{U_tX_tY_t Z_t}$ 
so that   
\beqno
& &q_{U_tX_tY_tZ_t}(u_t,x_t,y_t,z_t)
\\
&=&p_{U_tX_t}^{(\mu,\theta; q^{t-1}, \kappa^{t-1})}(u_t,x_t) W_1(y_1|x_t)W_2(z_t|y_t).
\eeqno
It is obvious that $q_t \in \hat{\cal P}_n(W_1,W_2)$ for 
$t=1,2, \cdots, n$. 
By (\ref{eqn:aasor}) and the above choice of $q_t$, we have 
\beqa
& &\Phi_{t,q^t, \kappa^t}^{(\mu,\theta)}
\nonumber\\
&=&\sum_{u_t,x_t,y_t,z_t} 
q_{U_t}(u_t)q_{X_t|U_t}(x_t|u_t)W_1(y_t|x_t)W_2(z_t|y_t)
\nonumber\\
& &\times 
\left\{
     \frac{W^\mu_1(y_t|x_t)}{q^{\mu}_{Y_t|U_t}(y_t|u_t)}
     \frac{p_{Z_t|V_t}(z_t|v_t)}{q_{Z_t}(z_t)}
\right\}^\theta
\nonumber\\
&=&
{\rm E}_{q_t}
\left[
\left\{
\frac{W_1^{\mu}(Y_t|X_t)}{q^{\mu}_{Y_t|U_t}(Y_t|U_t)}
\frac{p_{Z_t|U_t}(Z_t|V_t) }{q_{Z_t}(Z_t)}
\right\}^{\theta}
\right]
\nonumber\\
&=&
{\rm E}_{q_t}
\left[
\left\{
\frac{W_1^{\mu}(Y_t|X_t)}{q^{\mu}_{Y_t|U_t}(Y_t|U_t)}
\frac{q_{Z_t|U_t}(Z_t|U_t) }{q_{Z_t}(Z_t)}
\frac{p_{Z_t|V_t}(Z_t|V_t) }{q_{Z_t|U_t}(Z_t|U_t)}
\right\}^{\theta}
\right]
\nonumber\\
&\MLeq{a}&
\left(
{\rm E}_{q_t}
\left[
\left\{
\frac{W_1^{\mu}(Y_t|X_t)}{q^{\mu}_{Y_t|U_t}(Y_t|U_t)}
\frac{q_{Z_t|U_t}(Z_t|U_t) }{q_{Z_t}(Z_t)}
\right\}^{\frac{\theta}{1-\theta}}
\right]
\right)^{1-\theta}
\nonumber\\
& &\times \left(
{\rm E}_{q_t}
\left\{
\frac
{p_{Z_t|V_t}(Z_t|V_t)}
{q_{Z_t|U_t}(Z_t|U_t)}
\right\}\right)^{\theta}
\nonumber\\
&=&
\exp\left\{(1-\theta)
   \Omega^{(\mu,\frac{\theta}{1-\theta})}_{q_t}
   (X_tY_tZ_t|U_t)
\right\}
\nonumber\\
&\MEq{b}&
\exp\left\{\frac{1}{1+\lambda}\Omega^{(\mu,\lambda)}_{q_t}
   (X_tY_tZ_t|U_t)
\right\}
\nonumber\\
&\MLeq{c}&
\exp\left\{\frac{1}{1+\lambda}
   \hat{\Omega}_n^{(\mu,\lambda)}(W_1,W_2)
\right\}
\nonumber\\
&
\MEq{d}
&
\exp\left\{
\frac{1}{1+\lambda}
   {\Omega}^{(\mu,\lambda)}(W_1,W_2)
\right\}.
\label{eqn:sssto} 
\eeqa
Step (a) follows from H\"older's inequality. 
Step (b) follows from (\ref{eqn:abaddd}). 
Step (c) follows from 
$q_t \in \hat{\cal P}_n(W_1,W_2)$ and 
the definition of $\hat{\Omega}_n^{(\mu,\lambda)}$ $(W_1,W_2)$. 
Step (d) follows from Lemma \ref{lm:CardLmd} in Appendix \ref{sub:ApdaAAA}. 
To prove this lemma we bound the cardinality $|{\cal U}|$ 
appearing in the definition of 
$\hat{\Omega}_n^{(\mu,\lambda)}(W_1,W_2)$ 
to show that the bound $|{\cal U}|\leq |{\cal X}|$ 
is sufficient to describe $\hat{\Omega}_n^{(\mu,\lambda)}(W_1,W_2)$.
Hence we have the following: 
\beqa
& &\min_{\scs q^{n}\in {\cal Q}^n}
\frac{1}{n}\Omega^{(\mu,\theta)}_{p^{(n)}||q^{n}}(X^nY^nZ^n|{L_n})
\nonumber\\
&\leq  & \frac{1}{n}\Omega_{p^{(n)}||q^n}^{(\mu,\theta)}(X^nY^nZ^n|L_n)
 \MEq{a} \frac{1}{n}\sum_{t=1}^n \log 
\Phi_{t,q^t, \kappa^t}^{(\mu,\theta)}
\nonumber\\
&\MLeq{b}& \frac{1}{1+\lambda}
{\Omega}^{(\mu,\lambda)}(W_1,W_2).
\qquad \label{eqn:aQ1}
\eeqa
Step (a) follows from (\ref{eqn:defazzz}) 
in Lemma \ref{lm:keylm}.
Step (b) follows from (\ref{eqn:sssto}).
Since (\ref{eqn:aQ1}) holds for any ${n\geq 1}$ 
and any $p^{(n)}\in {\cal P}^{(n)}$ $(W_1,W_2)$, we have  
$$
\overline{\Omega}^{(\mu,\theta)}(W_1,W_2)
\leq 
\frac{1}{1+\gamma}
{\Omega}^{(\mu,\lambda)}(W_1,W_2).
$$
Thus, Proposition \ref{pro:mainpro} is proved.
\hfill \IEEEQED
}

{\it Proof of Theorem \ref{Th:main}: }
For $\theta \in (0,1)$, set 
\beq
\lambda=\frac{\theta}{1-\theta} 
\Leftrightarrow \theta=\frac{\lambda}{1+\lambda}. 
\label{eqn:abadd}
\eeq
Then we have the following:
\beqno
& &G(R_1,R_2|W_1,W_2)
\nonumber\\
&\MGeq{a}& 
\frac{
\theta(\mu R_1 + R_2)
-\overline{\Omega}^{(\mu,\theta)}(W_1,W_2)
}
{1+\theta(1+\mu)}
\nonumber\\
&\MGeq{b}& 
\frac{\frac{\lambda}{1+\lambda}(\mu R_1 + R_2)
-\frac{1}{1+\lambda}{\Omega}^{(\mu,\lambda)}(W_1,W_2)
}
{1+\frac{\lambda}{1+\lambda}(1+\mu)}
\nonumber\\
&=&
\frac{\lambda(\mu R_1 + R_2)-\Omega^{(\mu,\lambda)}(W_1,W_2)
}
{1+\lambda+\lambda(1+\mu)}
\\
&=&F^{(\mu,\lambda)}(\mu R_1+R_2|W_1,W_2). 
\label{eqn:Zddd}
\eeqno
Step (a) follows from Corollary \ref{co:corOne}. Step (b) follows from 
Proposition \ref{pro:mainpro} and (\ref{eqn:abadd}). 
Since (\ref{eqn:Zddd}) holds for any $\lambda,\mu>0$, we have (\ref{eqn:mainIeq}) 
in Theorem \ref{Th:main}. 
\hfill\IEEEQED

\newcommand{\ApdaAAAb}{

We have the following lemma.
\begin{lm} \label{lm:CardLmd} 
For each integer $n \geq 2$, we define 
\beqno
&      &\hat{\Omega}_n^{(\mu,\lambda)}(W_1,W_2)
\\
&\defeq& \max_{\scs q=q_{UXYZ}: 
         U\leftrightarrow X \leftrightarrow Y \leftrightarrow Z,
        \atop{\scs
           \atop{\scs
           q_{Y|X}=W_1,q_{Z|Y}=W_2,
            \atop{\scs  
            |{\cal U}|\leq |{\cal L}_n||{\cal Y}|^{n-1}  
     }}}}
\\
& &\Omega^{(\mu,\lambda)}(W_1,W_2)
\\
&\defeq& \max_{\scs q=q_{UXYZ}: U\leftrightarrow X \leftrightarrow Y \leftrightarrow Z,
        \atop{\scs
           \atop{\scs
           q_{Y|X}=W_1,q_{Z|Y}=W_2,
            \atop{\scs  
            |{\cal U}|\leq |{\cal X}|
     }}}}
\Omega^{(\mu,\lambda)}_{q}(XYZ|U).
\eeqno
Then we have 
\beqno
& &\hat{\Omega}^{(\mu,\lambda)}(W_1,W_2)
  =\Omega^{(\mu,\lambda)}(W_1,W_2).
\eeqno

\end{lm}

{\it Proof:} We bound  the cardinality $|{\cal U}|$ of ${U}$ 
to show that the bound $|{\cal U}| \leq |{\cal X}|$
is sufficient to describe 
$\hat{\Omega}_n^{(\mu,\lambda)}$ 
$(W_1,W_2)$ and 
$\tilde{\Omega}_n^{(\mu,\lambda)}$ 
$(W_1,W_2)$. 
Observe that 
\beqa
\hspace*{-5mm}& &q_{{X}}(x)
=\sum_{u\in {\cal U}}q_U(u)
q_{{X}|U}(x|u),
\label{eqn:asdf}
\\
\hspace*{-5mm}
& &\Lambda_{q}^{(\mu,\lambda)}(XYZ|U)
=\sum_{u\in {\cal U}}q_U(u)
\zeta^{(\mu, \lambda)}(q_{{X}|U}(\cdot|u)),
\label{eqn:aqqqa}
\eeqa
where 
\beqno
& &\zeta^{(\mu, \lambda)}(q_{{X}|U}(\cdot|u))
\\
&\defeq & \sum_{(x,y,z)\in{\cal X}\times{\cal Y}\times{\cal Z}}
q_{{X}|U}(x|u)W_1(y|x)W_2(z|y)
\\
& &\times \exp\left\{\lambda \omega^{(\mu)}_{q}(x,y,z|u)\right\}
\eeqno
For the quantities $q_{Z}(\cdot)$ 
contained in the forms of 
$\zeta^{(\mu,\lambda)}$ 
$(q_{X|U}(\cdot|u)),$ $u\in {\cal U}$, we regard them as 
constants under (\ref{eqn:asdf}). For each $u \in{\cal U}$, 
$\zeta^{(\mu, \lambda)}(q_{{X}|U}(\cdot|u))$
are continuous functions of $q_{{X}|U}(\cdot|u)$.
Then by the support lemma, 
$$
|{\cal U}| \leq |{\cal X}|-1 +1= |{\cal X}| 
$$
is sufficient to express $|{\cal X}|-1$ 
values of (\ref{eqn:asdf}) 
and one value of (\ref{eqn:aqqqa}). 
\hfill \IEEEQED
}

\section{Concluding Remarks}

For the DBC, we have derived an explicit lower bound of the optimal 
exponent function on the correct probability of decoding for rates 
outside the capacity region. Our method for the DBC can also be applied 
to the derivation of an explicit lower bound of 
the optimal exponent function outside the capacity region 
for the asymmetric broadcast channels(ABCs)( or said the broadcast 
channels with degraded message sets) investigated by \cite{ck},
\cite{km77}-\cite{KaMer11}. In fact the author \cite{OohamaAbcSita13} 
succeeded deriving an explicit 
lower bound of the exponent function that is positive for rates outside 
the capacity region of the ABC. In the case of ABC, some additional 
techniques are also needed. 

%
%
%
%
%

\section*{\empty}
\appendix

\ApdaAAA
\ApdaAAAb

\ApdaRegAAA
\ApdaAAC

\Apda
\ApdLemmaA
\ApdLemmaB
\ApdLemmaC
\Apdc
\ApdLemmaCContA
\ApdLemmaCContB


\begin{thebibliography}{99}

\bibitem{cov72}T.~M. Cover, ``Broadcast channels,'' 
{\em IEEE Trans. Inform. Theory}, vol. IT-18, 
no.1, pp.~2--13, Jan. 1972.

\bibitem{bgm73}P. P. Bergmans, ``Random coding theorems 
for broadcast channels with degraded components," 
{\it IEEE Trans. Inform. Theory}, vol. IT-19, pp. 197-207, 
Mar. 1973.


\bibitem{gal74}
R. G. Gallager, ``Capacity and coding for degraded broadcast channels," 
{\it Problemy Peredachi Informatsii}, vol. 10, pp. 3-14, 
July-Sept. 1974.

\bibitem{ak75}
R. F. Ahlswede and J. K\"orner, ``Source coding with side information
and a converse for degraded broadcast channels," 
{\it IEEE Trans. Inform. Theory}, vol. IT-21, pp. 629-637, Nov. 1975.



\bibitem{agk76}R. Ahlswede, P. G\`as, and J. K\"orner, 
``Bounds on conditional probabilities with applications 
in multi-user communication," {\it Z. Wahrscheinlichkeitstheorie 
verw. Gebiete}, vol. 34, pp. 157-177, 1976.







\bibitem{ck} 
I. Csisz\'ar and J. K\"orner, 
{\it Information Theory: Coding Theorems for Discrete 
Memoryless Systems.} Academic Press, New York, 1981.





\bibitem{ari}S. Arimoto,
``On the converse to the coding theorem for discrete memoryless 
channels,'' {\it IEEE Trans. Inform. Theory,} vol. IT-19, no. 3, pp. 
357-359, May 1973.

\bibitem{dk}G. Dueck and J. K\"orner, ``Reliability function 
of a discrete memoryless channel at rates above capacity,'' 
{\it IEEE Trans. Inform. Theory,} vol. IT-25, no. 1, pp. 82--85, Jan. 1979.

\bibitem{han}
T. S. Han, {\it Information-Spectrum Methods in Information
Theory. }Springer-Verlag, Berlin, New York, 2002. The Japanese 
edition was published by Baifukan-publisher, Tokyo, 1998.

\bibitem{km77}J. K\"orner and K. Marton, 
``General broadcast channels with degraded message sets,"
{\it IEEE Trans. Inform. Theory,} vol. IT-23, no. 1, 
pp. 60-64, Jan 1977. 

\bibitem{ks80}J. K\"orner and A. Sgarro, 
``Universally attainable error exponents for broadcast 
channels with degraded message sets,"
{\it IEEE Trans. Inform. Theory,} vol. IT-26, no. 6, 
pp. 670-679, Nov. 1980.

\bibitem{KaMer11}Y. Kaspi and N. Merhav, 
``Error exponents for broadcast channels with 
degraded Message sets,"
{\it IEEE Trans. Inform. Theory,} vol. 57, no. 1, 
pp.101-123, Jan. 2011.

\bibitem{OohamaAbcSita13}Y. Oohama, 
``New converse for asymmetric broadcast channels,'' 
{\it Proceedings of the 36th Symposium on Information Theory 
and its Applications(SITA2013)}, pp. 273--278, Ito, Shizuoka, 
Japan, Nov. 26-29, 2013.

\end{thebibliography}
\end{document}